\documentclass[fleqn,usenatbib]{mnras}

\usepackage{newtxtext,newtxmath}
\usepackage[T1]{fontenc}

\DeclareRobustCommand{\VAN}[3]{#2}
\let\VANthebibliography\thebibliography
\def\thebibliography{\DeclareRobustCommand{\VAN}[3]{##3}\VANthebibliography}

\usepackage{graphicx}
\usepackage{amsmath}
\usepackage{longtable}
\usepackage{subfig}

\title[MOSDEF $z\sim2$ Completeness]{The MOSDEF Survey: Towards a Complete Census of the \textit{z} $\sim$ 2.3 Star-forming Galaxy Population$^{1}$}

\author[J. N. Runco et al.]{Jordan N. Runco,$^{2}$\thanks{E-mail: jrunco@astro.ucla.edu}
Alice E. Shapley,$^{2}$
Ryan L. Sanders,$^{3,4}$
Mariska Kriek,$^{5,6}$\newauthor
Naveen A. Reddy,$^{7}$
Alison L. Coil,$^{8}$
Bahram Mobasher,$^{7}$
Brian Siana,$^{7}$\newauthor
Michael W. Topping,$^{2,9}$
William R. Freeman,$^{7}$
Irene Shivaei,$^{9}$
Mojegan Azadi,$^{10}$\newauthor
Sedona H. Price,$^{11}$
Gene C. K. Leung,$^{12}$
Tara Fetherolf,$^{7}$
Laura de Groot,$^{13}$\newauthor
Tom Zick,$^{6}$
Francesca M. Fornasini,$^{14}$
Guillermo Barro$^{15}$
\\
$^{1}$Based on data obtained at the W.M. Keck Observatory, which is operated as a scientific partnership among the California Institute of \\ Technology, the University of California,  and the National Aeronautics and Space Administration, and was made possible by the generous  \\ financial support  of the W.M. Keck Foundation.\\
$^{2}$Physics \& Astronomy Department, University of California: Los Angeles, 430 Portola Plaza, Los Angeles, CA 90095, USA\\
$^{3}$Department of Physics, University of California, Davis, One Shields Ave, Davis, CA 95616, USA\\
$^{4}$Hubble Fellow\\
$^{5}$Leiden Observatory, Leiden University, PO Box 9513, NL-2300 RA Leiden, The Netherlands \\
$^{6}$Astronomy Department, University of California, Berkeley, CA 94720, USA\\
$^{7}$Department of Physics \& Astronomy, University of California, Riverside, 900 University Avenue, Riverside, CA 92521, USA\\
$^{8}$Center for Astrophysics and Space Sciences, University of California, San Diego, 9500 Gilman Dr., La Jolla, CA 92093-0424, USA\\
$^{9}$Department of Astronomy/Steward Observatory, 933 North Cherry Ave, Rm N204, Tucson, AZ 85721-0065, USA\\
$^{10}$Harvard-Smithsonian Center for Astrophysics, 60 Garden Street, Cambridge, MA, 02138, USA \\
$^{11}$Max-Planck-Institut f\"ur Extraterrestrische Physik, Postfach 1312, Garching, 85741, Germany \\
$^{12}$Department of Astronomy, The University of Texas at Austin, 2515 Speedway Blvd Stop C1400, Austin, TX 78712, USA\\
$^{13}$Department of Physics, The College of Wooster, 1189 Beall Avenue, Wooster, OH 44691, USA\\
$^{14}$Department of Physics and Astronomy, Stonehill College, 320 Washington Street, Easton, MA 02357, USA \\
$^{15}$Department of Physics, University of the Pacific, 3601 Pacific Ave, Stockton, CA 95211, USA \\
}

\date{Accepted XXX. Received YYY; in original form ZZZ}

\pubyear{2022}

\begin{document}
\label{firstpage}
\pagerange{\pageref{firstpage}--\pageref{lastpage}}
\maketitle

\begin{abstract}
We analyze the completeness of the MOSDEF survey, in which $z\sim2$ galaxies were selected for rest-optical spectroscopy from well-studied \textit{HST} extragalactic legacy fields down to a fixed rest-optical magnitude limit ($H_{\rm{AB}} =$ 24.5). The subset of $z\sim2$ MOSDEF galaxies with high signal-to-noise (S/N) emission-line detections analyzed in previous work represents a small minority ($<10$\%) of possible $z\sim2$ MOSDEF targets. It is therefore crucial to understand how representative this high S/N subsample is, while also more fully exploiting the MOSDEF spectroscopic sample. Using spectral-energy-distribution (SED) models and rest-optical spectral stacking, we compare the MOSDEF $z\sim2$ high S/N subsample with the full MOSDEF sample of $z\sim2$ star-forming galaxies with redshifts, the latter representing an increase in sample size of more than a factor of three. We find that both samples have similar emission-line properties, in particular in terms of the magnitude of the offset from the local star-forming sequence on the [N~\textsc{II}] BPT diagram. There are small differences in median host galaxy properties, including the stellar mass ($M_{\ast}$), star-formation rate (SFR) and specific SFR (sSFR), and UVJ colors; however, these offsets are minor considering the wide spread of the distributions. Using SED modeling, we also demonstrate that the sample of $z\sim2$ star-forming galaxies observed by the MOSDEF survey is representative of the parent catalog of available such targets. We conclude that previous MOSDEF results on the evolution of star-forming galaxy emission-line properties were unbiased relative to the parent $z\sim2$ galaxy population.
\end{abstract}


\section{Introduction} \label{sec:intro}

One of the most powerful tools for studying galaxies across time is rest-frame optical emission-line spectroscopy. Such measurements  provide information about the properties of a galaxy, including its active galactic nucleus (AGN) activity, star-formation rate (SFR), virial and non-virial dynamics (e.g., outflows), dust extinction, and properties of the ionized interstellar medium (ISM) such as the electron density ($n_{\rm{e}}$), metallicity, and ionization parameter ($U$; i.e., the ratio of ionizing photon density to hydrogen, and therefore $n_{\rm{e}}$). Applying this tool to galaxies observed during the peak epoch of star-formation in the universe ($z\sim 2$) is especially effective for understanding the origin of well-known patterns exhibited by galaxies in the universe today.

Over the past decade, the deployment of multi-object near-infrared (IR) spectrographs on large ground-based telescopes --- e.g., the MultiObject Spectrometer For Infra-Red Exploration (MOSFIRE; \citealt{mcl12}) on the Keck I telescope --- has enabled the collection of rest-optical spectra for large statistical samples of high-redshift ($z\sim1.5-3.5$) galaxies. One survey utilizing MOSFIRE is the MOSFIRE Deep Evolution Field (MOSDEF; \citealt{kri15}) survey, which contains $\sim$1500 galaxies with Keck/MOSFIRE at $1.4 \lesssim z \lesssim 3.8$ with roughly half at $z\sim2$. 

One of the goals of the MOSDEF sample was to target and study a roughly stellar-mass-complete sample at high redshift. However, there are many stages in which incompleteness can be introduced and thus complicate the achievement of this goal. The initial parent catalog of MOSDEF galaxies was composed of galaxies with estimated redshifts within fixed targeted ranges to optimize the detection of strong rest-optical emission lines, and brighter than a fixed $H_{\rm{AB}}$ magnitude limit. However, there is targeting incompleteness between this parent catalog and the actual sample of MOSDEF galaxies observed. Furthermore, there is spectroscopic incompleteness, between the sample of MOSDEF galaxies observed and the one for which MOSFIRE spectroscopic redshifts were measured. Finally, there is detection incompleteness, between the MOSDEF sample with spectroscopic redshifts, and the sample for which multiple rest-optical emission lines are detected and analyzed in previous works \citep[e.g.,][]{sha15,san18,shi20,top20a,run21}.  

Notably, a recent comparison between the MOSDEF and KBSS-MOSFIRE $z\sim2$ surveys revealed that, for MOSDEF, the subset of star-forming galaxies with S/N $\geq$ 3 for H$\beta$ and H$\alpha$ display different physical properties from those of the entire catalog of galaxies observed at $z\sim2$ \citep{run22}. Specifically, the $z\sim2$ subsample of galaxies with high S/N spectra have a lower median stellar mass ($M_{\ast}$) and stellar population age, a higher median star formation rate (SFR) and specific star formation rate (sSFR), and a bluer median U$-$V color compared to the $z\sim2$ observed sample from which it was drawn. \citet{run22} showed that this high S/N subsample is incomplete with respect to red, massive galaxies targeted by MOSDEF with older and less intense star formation. Accordingly, this subsample does not fully represent the complete demographics of $z\sim2$ galaxy sample with attempted MOSDEF spectroscopic observations. While the MOSDEF incompleteness appears most severe for galaxies that are not actively forming stars, it is still a matter of concern how representative the $z\sim2$ subsample of MOSDEF galaxies with high S/N spectra is of the overall $z\sim2$ star-forming galaxy population.

There has been much previous work investigating the rest-optical emission-line properties of the $z\sim2$ MOSDEF sample, implementing selection criteria to isolate subsets of the $z\sim2$ MOSDEF sample with high S/N to obtain clean results unobstructed by low S/N spectra (e.g., \citealt{sha15, sha19, san16, san18, san20, san21, jeo20, top20a, top20b, run21, run22}). It is important to note that the subsample of MOSDEF galaxies with high S/N analyzed in, e.g., \citet{run22} and \citet{san18} comprises only $\sim$32\% (250/786 galaxies) of the full $z\sim2$ sample observed by MOSDEF. Furthermore, even when limited to the set of star-forming galaxies alone --- since it is these star-forming galaxies we seek to understand using emission-line diagrams  --- the high S/N subsample is still only $\sim$41\% (250/617 galaxies) of the full $z\sim2$ observed sample. The question remains if the differences in the physical properties between the galaxies with and without high S/N correspond to a fundamental difference in emission-line properties as well. 

Here we aim to investigate this question using spectral stacking, regardless of the emission lines detected. We construct a significantly larger sample of $\sim500$ MOSDEF star-forming galaxies, where the only requirement for inclusion is the measurement of a spectroscopic redshift, and rest-frame optical (UVJ) colors corresponding to a star-forming (not quiescent) spectral-energy-distribution (SED). This sample, which we refer to as the ``$z\sim2$ stacked sample" represents a much more complete portion of the star-forming galaxies observed by the MOSDEF survey. Previous MOSDEF studies (e.g., \citealt{san21}) have utilized spectral stacking; however, these studies only incorporate a minority of the MOSDEF $z\sim2$ star-forming sample that has at least one emission-line detection.

We perform a comparison of the locations of the larger, more complete $z\sim2$ stacked sample and the previously studied, $z\sim2$ high-S/N subsample on the [O~\textsc{III}]$\lambda$5008/H$\beta$ vs. [N~\textsc{II}]$\lambda$6585/H$\alpha$ diagram (first introduced by \citealt{bal81} and commonly referred to as the ``[N~\textsc{II}] BPT diagram''), and the [O~\textsc{III}]$\lambda$5008/H$\beta$ vs. [S~\textsc{II}]$\lambda\lambda$6718,6733/H$\alpha$ diagram (first introduced in \citealt{vei87} and commonly referred to as the ``[S~\textsc{II}] BPT diagram''). These diagrams can be used to infer whether the dominant source of ionizing radiation in an emission-line galaxy is an AGN or star formation, given the distinct regions in emission-line ratio space occupied by AGNs and star-forming galaxies. Another diagram frequently used to characterize emission-line galaxies is the [O~\textsc{III}]$\lambda\lambda$4960,5008/[O~\textsc{II}]$\lambda\lambda$3727,3730 (O$_{32}$) vs. ([O~\textsc{III}]$\lambda\lambda$4960,5008+[O~\textsc{II}]$\lambda\lambda$3727,3730)/H$\beta$ (R$_{23}$) diagram. O$_{32}$ (R$_{23}$) roughly probes the ionization parameter (metallicity) of star-forming galaxies (e.g., \citealt{lil03, nak13}). It has been shown that metallicity increases from high excitation (high O$_{32}$ \& R$_{23}$) to low excitation (low O$_{32}$ \& R$_{23}$) \citep{and13, sha15}. 

Gaining a complete understanding of the [N~\textsc{II}] BPT diagram is essential because rest-optical emission-lines are used as calibrations for many physical and chemical galaxy properties not directly observable at high redshift. Many such calibrations exist for star-forming galaxies in the local universe (e.g., strength of emission-line ratios such as [N~\textsc{II}]$\lambda$6585/H$\alpha$ correlate with gas-phase oxygen abundance \citealt{pet04}). However, it is well documented that star-forming galaxies at $z>1$ show elevated [N~\textsc{II}]$\lambda$6585/H$\alpha$ at fixed [O~\textsc{III}]$\lambda$5008/H$\beta$ (or vice versa; e.g. \citealt{sha05, sha15, erb06b, liu08, ste14, str17, run21}) on average compared to local $z\sim0$ galaxies in the Sloan Digital Sky Survey (SDSS; \citealt{yor00}). Numerous explanations for this systematic offset have been suggested, including galaxy selection effects, shocks, unresolved AGN activity (i.e., galaxies with star-formation and AGN activity is mistaken for only star-formation activity), gas-phase N/O abundance ratio differences, and variations in physical properties of H~\textsc{II} regions inside galaxies such as ionization parameter, electron densities, density structure, and the hardness of the ionizing spectra at fixed metallicities (e.g., \citealt{liu08, bri08, wri10, kew13, yeh13, jun14, mas14, ste14, ste16, coi15, sha15, sha19, san16, str17, str18, fre19, kas19, top20a, run21}). 

Current MOSDEF results favor the latter idea (i.e., that $z\sim2$ star-forming galaxies contain a harder ionizing spectrum at fixed nebular oxygen abundance compared to the population of star-forming galaxies at $z\sim0$) being the main driver of the observed [N~\textsc{II}] BPT offset. This difference in the ionizing spectrum arises due to $\alpha$-enhancement in the massive stars of $z\sim2$ star-forming galaxies \citep{sha19, top20a, red21, run21}. Studies using the Keck Baryonic Structure Survey (KBSS: \citealt{ste14}), agree with this interpretation of the offset (e.g., \citealt{ste14, ste16, str17}). However, without a complete understanding of the $z>1$ [N~\textsc{II}] BPT offset, it is unclear how established calibrations between line ratios and physical properties must be modified to be valid for high-redshift galaxies.

Therefore, this study will also analyze how emission-line ratios commonly used as metallicity calibrators (e.g., [N~\textsc{II}]$\lambda$6585/H$\alpha$; \citealt{pet04}) and dust attenuation tracers (H$\alpha$/H$\beta$; e.g., \citealt{kas13, sha22}) correlate with stellar mass. We will investigate potential offsets between the full catalog of MOSDEF $z\sim2$ star-forming galaxies using spectral stacking in comparison with the subset of star-forming galaxies that have high S/N $z\sim2$ spectra. The galaxy properties of these different MOSDEF samples will be estimated using broadband SED fitting to correlate with any offset in the emission-line properties. In addition, we will track the offset between the multiple MOSDEF samples with the local $z\sim0$ SDSS star-forming sequence to quantify evolution in emission-line properties over the past $\sim$10 Gyr. 

Finally, the 786 $z\sim 2$ galaxies observed as part of MOSDEF do not comprise the full sample satisfying the simple MOSDEF selection criteria  --- i.e., the parent catalog of 3,780 galaxies in \textit{HST} extragalactic legacy fields  with estimated redshifts within the $z\sim 2$ range targeted by MOSDEF and $H$-band (rest-optical) apparent magnitudes down to a fixed limit ($H_{\rm AB}=24.5$).  Therefore, in this study we also utilize broadband SED fitting and compare the full parent catalog of available $z\sim 2$ galaxies with the subset actually observed by MOSDEF. In this part of our analysis, we aim to understand how representative the MOSDEF observed sample is. 

Section \ref{sec:sample_selection} details the MOSDEF survey, defines the multiple MOSDEF samples investigated in this study, and the methodology for the SED modeling used to estimate the physical properties of the galaxies and spectra stacking. 
Section \ref{sec:results} presents the results of the MOSDEF sample comparisons using the spectral stacking and SED fitting techniques, 
while Section \ref{sec:discussion} provides a discussion on how these results relate to past MOSDEF studies. 
Finally, Section \ref{sec:summary} summarizes the key results and looks ahead to future work. 
We adopt the following abbreviations for emission-line ratios used frequently throughout the paper.
\begin{equation}
\label{eqn:N2_abbreviation}
    \rm{N2 = [N~\textsc{II}]\lambda6585/H\alpha}
\end{equation}
\begin{equation}
\label{eqn:S2_abbreviation}
    \rm{S2 = [S~\textsc{II}]\lambda\lambda6718,6733/H\alpha}
\end{equation}
\begin{equation}
\label{eqn:O3_abbreviation}
    \rm{O3 = [O~\textsc{III}]\lambda5008/H\beta}
\end{equation}
\begin{equation}
\label{eqn:O3N2_abbreviation}
    \rm{O3N2 = O3/N2}
\end{equation}
\begin{equation}
\label{eqn:O3S2_abbreviation}
    \rm{O3S2 = O3/S2}
\end{equation}
\begin{equation}
\label{eqn:O32_abbreviation}
    \rm{O_{32} = [O~\textsc{III}]\lambda\lambda4960,5008/[O~\textsc{II}]\lambda\lambda3727,3730}
\end{equation}
\begin{equation}
\label{eqn:R23_abbreviation}
    \rm{R_{23} = ([O~\textsc{III}]\lambda\lambda4960,5008+[O~\textsc{II}]\lambda\lambda3727,3730)/H\beta}
\end{equation}
Throughout this paper, all emission-line wavelengths are in vacuum, and we adopt a $\Lambda$-CDM cosmology with $H_0$ = 70 km s$^{-1}$ Mpc$^{-1}$, $\Omega_{\rm{m}}$ = 0.3, and $\Omega_\Lambda$ = 0.7.

\section{Observations, Sample Selection, \& Methods} \label{sec:sample_selection}

Here we provide an overview of the MOSDEF survey and describe the selection methods for the various samples used in this study (Section \ref{subsec:mosdef_sample_stack_paper}). 
Additionally, we discuss our approach to SED fitting and estimating key galaxy properties (Section \ref{subsec:sed_fitting}), the methodologies for both emission-line fitting (Section \ref{subsec:emline_fits}) and stacking spectra (Section \ref{subsec:spectral_stacking}), and the selection methods for the local $z\sim0$ SDSS comparison sample \ref{subsec:local_sdss_sample_stacking}.

\begin{figure*}
    \includegraphics[width=0.49\linewidth]{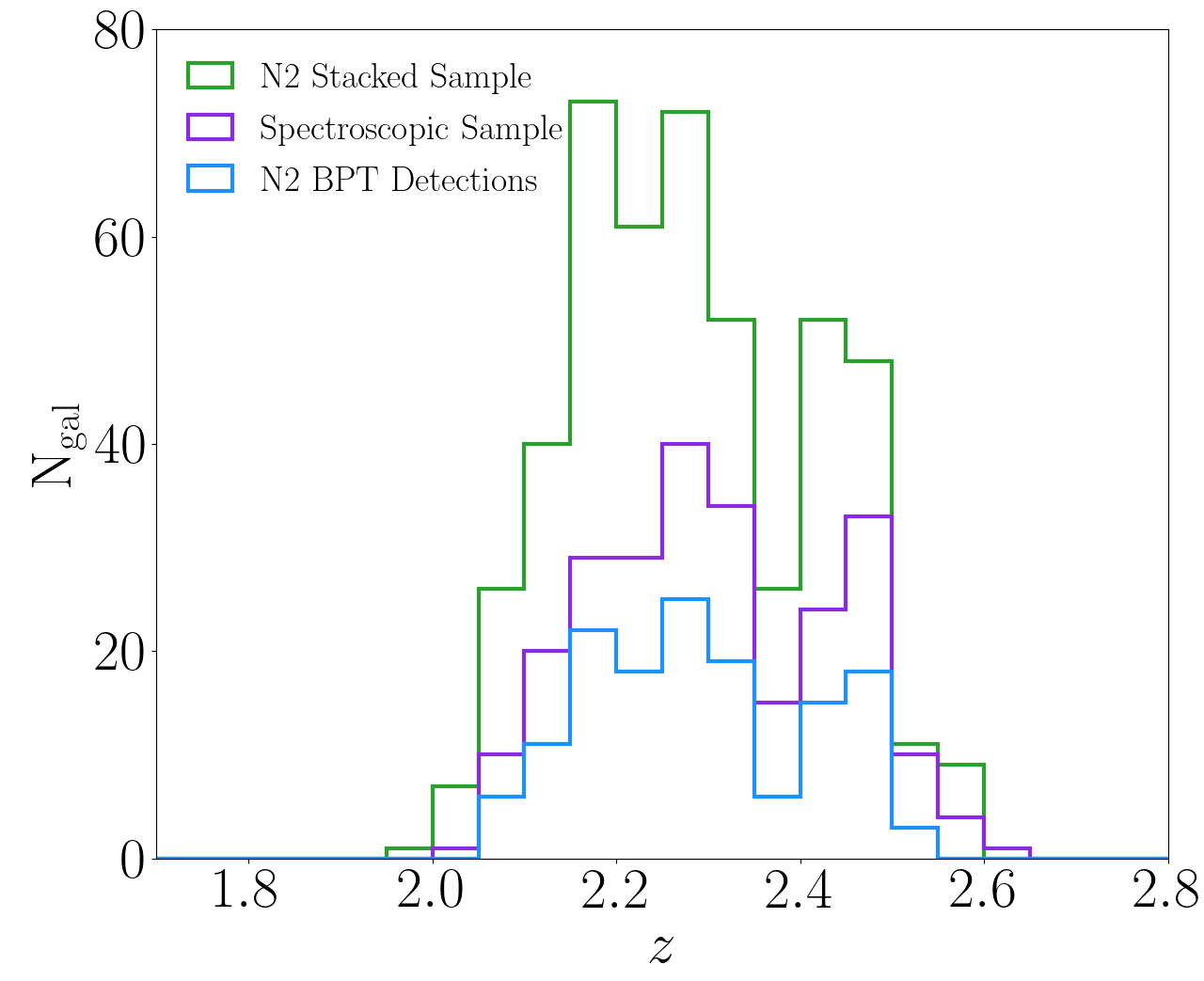}
    \includegraphics[width=0.49\linewidth]{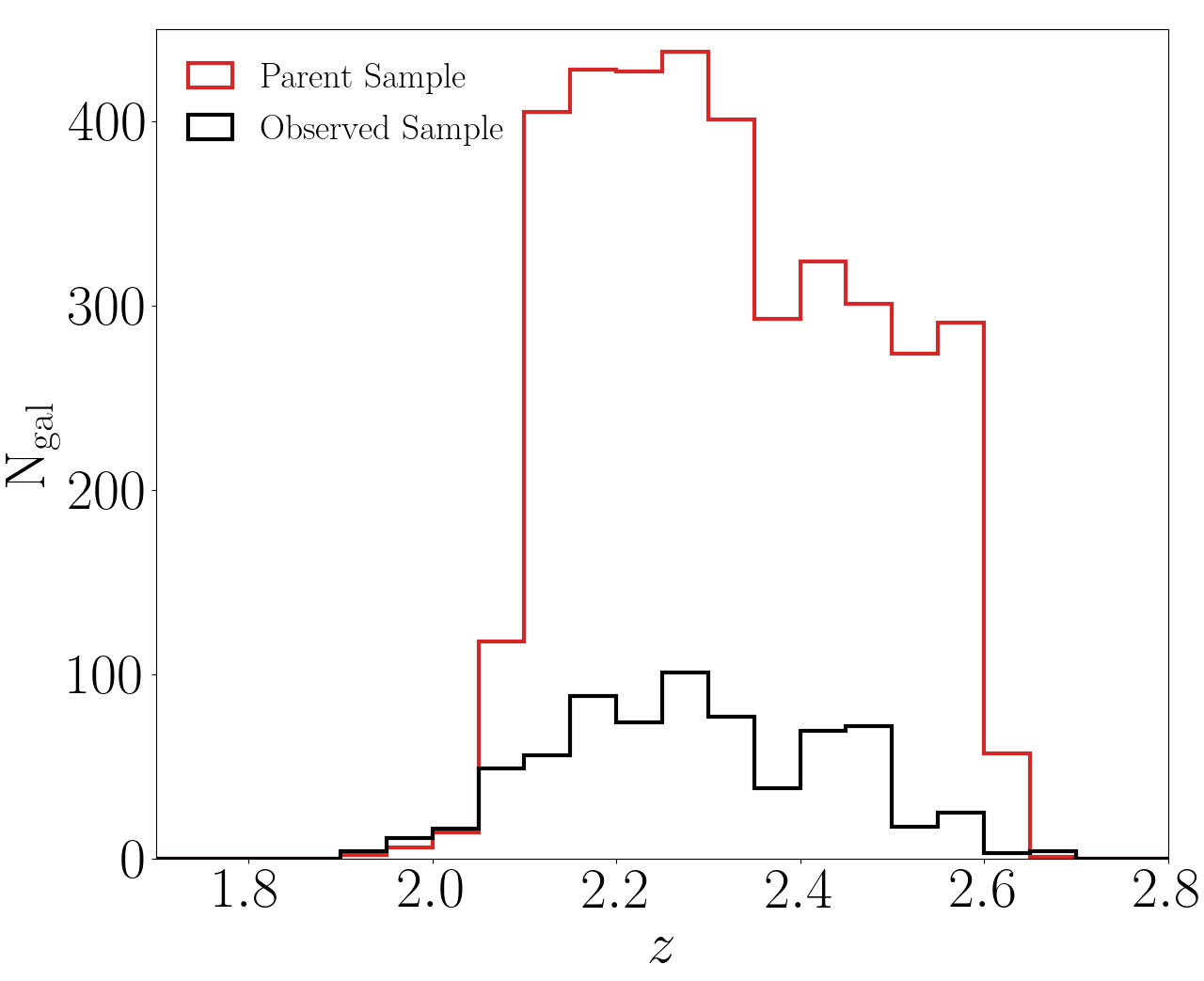}
    \caption{Redshift distributions for the different MOSDEF samples analyzed in this study. The left panel shows the MOSDEF $z\sim2$ N2 stacked sample (green; $z_{\rm{med}}=2.28$), $z\sim2$ spectroscopic sample (purple; $z_{\rm{med}}=2.29$), and $z\sim2$ N2 BPT detection sample (blue; $z_{\rm{med}}=2.29$). The right panel shows the MOSDEF $z\sim2$ observed sample (black; $z_{\rm{med}}=2.29$) and $z\sim2$ parent sample (red; $z_{\rm{med}}=2.31$).}
    \label{fig:redshift_hists_stacks}
\end{figure*}

\subsection{The MOSDEF Survey} \label{subsec:mosdef_sample_stack_paper}

In the MOSDEF survey, galaxies were targeted within five well-studied \textit{HST} extragalactic legacy fields covered by the CANDELS and 3D-HST surveys: AEGIS, COSMOS, GOODS-N, GOODS-S, and UDS \citep{gro11, koe11, mom16}. These fields were selected because of the large amounts of ancillary data available. Multi-wavelength photometric observations enable us to perform robust SED fitting (see below), and observations outside the rest-optical (i.e., X-ray and mid-IR) provide useful information on the incidence of AGN.
Within these fields, MOSDEF targets galaxies in three redshift ranges: $1.37 \leq z \leq 1.70$, $2.09 \leq z \leq 2.61$, and $2.95 \leq z \leq 3.80$. These redshift ranges were selected to optimize the detection of strong rest-frame optical emission-lines (e.g., [O~\textsc{II}]$\lambda\lambda$3727,3730, H$\beta$, [O~\textsc{III}]$\lambda\lambda$4960,5008, H$\alpha$, [N~\textsc{II}]$\lambda$6585, and [S~\textsc{II}]$\lambda\lambda$6718,6733) within windows of atmospheric transmission.
Additionally, galaxies were selected based on $H$-band (F160W) magnitude, with brightness limits of $H_{\rm{AB}} =$ 24.0, 24.5, and 25.0, respectively, for the lowest, middle and highest redshift ranges.
With these redshift and rest-optical brightness selection cuts, there are 10632 galaxies available for MOSDEF to observe between $z\sim1.3-3.8$, with 3780 galaxies at $z\sim2$. Hereafter, we refer to these 3780 galaxies at $z\sim2$ as the MOSDEF ``$z\sim2$ parent sample.''

MOSDEF was awarded 48.5 Keck/MOSFIRE nights between 2012-2016.
Despite this large allocation, it was not possible to observe the full set of 10632 galaxies that meet the MOSDEF selection criteria. Approximately 1500 galaxies were targeted over the three redshift bins. These galaxies were targeted primarily based on location in the sky for optimization of the MOSFIRE slit masks. About half of the targeted galaxies (786) fall in the middle $z\sim2$ redshift bin. Of these 786 galaxies, 77 are serendipitous detections and not actually targeted by MOSDEF. 
For a full description of MOSDEF observing details, see \citet{kri15}.

In \citet{run22}, we identified two MOSDEF samples. The MOSDEF ``$z\sim2$ targeted sample'' contained all 786 galaxies observed by MOSDEF at $1.9 \leq z \leq 2.7$ (in this study we more accurately refer to it as the ``$z\sim2$ observed sample'' because it contains 77 serendipitously observed galaxies), while the MOSDEF ``$z\sim2$ spectroscopic sample'' was based on the sample analyzed by \citet{san18}. The sample from \citet{san18} comprised 260 star-forming galaxies (i.e., no AGN) with S/N$_{\rm{H\alpha}}$ and S/N$_{\rm{H\beta}}$ $\geq$ 3. Additionally, we removed galaxies with log$_{10}(M_{\ast}/M_{\odot}) < 9.0$ due to the incompleteness of the MOSDEF survey at low mass (see \citealt{shi15} for details). 
AGN were removed based on X-ray luminosity, IR colors, or if N2 $\geq 0.5$ \citep{coi15, aza17, leu17}. 
In \citet{run22}, we removed 9/260 galaxies due to updated SED fitting yielding a $M_{\ast}$ estimate below the 10$^{9}$ $M_{\odot}$ cutoff, and one more additional galaxy due to it not being in the 3D-HST v4.1 catalog which is needed for the SED fitting. Therefore, the final MOSDEF ``$z\sim2$ spectroscopic sample'' from \citet{run22} contained 250 star-forming galaxies with S/N$_{\rm{H\alpha}}$ and S/N$_{\rm{H\beta}}$ $\geq$ 3 and log$_{10}(M_{\ast}/M_{\odot}) \geq 9.0$, which is featured in this work.

Note that there are 83 galaxies in the $z\sim2$ observed sample that are not in the $z\sim2$ parent sample. Of these 83 galaxies, 44 fell serendipitously in MOSDEF slits and were either fainter than the MOSDEF $H$-band magnitude limit, or else had estimated redshifts slightly outside the MOSDEF target range. The remaining 39 galaxies met the selection criteria in the 3D-HST v2.1 or v4.0 catalogs, which were used to select galaxies for observations during the early stages of the MOSDEF survey, before the finalized 3D-HST v4.1 catalogs were available. However, in the updated v4.1 catalogs used to obtain the $z\sim2$ parent sample in this study, changes to either the photometry or the redshift caused these galaxies no longer to satisfy  the MOSDEF selection criteria. For example, updated photometry measurements could lead to galaxies falling below the $H_{\rm{AB}} =$ 24.5 cut applied to the v4.1 catalog, if they were near that threshold in the v2.1 catalog. In terms of redshift, MOSDEF prioritized different redshift sources in the following order for targeting purposes: previous spectroscopic redshifts, grism redshifts, and then photometric redshifts. It is possible that the most reliable redshift in the v2.1 catalog was in the $z=2.09-2.61$ range, but the updated most reliable redshift in the v4.1 catalog was not (e.g., the most robust redshift in the v2.1 catalog was a photometric redshift at $2.09\leq z\leq 2.61$, and an updated grism redshift in the v4.1 catalog fell outside of this range). 

There is a large difference between the $z\sim2$ spectroscopic and $z\sim2$ observed samples, as the former only accounts for $\sim$32\% of the sample from which it was drawn. And there is one final layer of incompleteness that must be considered. Specifically, of the 250 galaxies in the $z\sim2$ spectroscopic sample, only 143, 155, and 181 had S/N $\geq$ 3 for all of the necessary lines to be analyzed on the [N~\textsc{II}] BPT, [S~\textsc{II}] BPT, and O$_{32}$ vs. R$_{23}$ diagram, respectively, which is only $\sim$18\%, $\sim$20\%, and $\sim$23\%, respectively, of the $z\sim2$ observed sample. In this study, we will refer to these 143, 156, and 181 galaxy samples, respectively, as the ``$z\sim2$ N2 BPT detection sample", ``$z\sim2$ S2 BPT detection sample", and ``$z\sim2$ O$_{32}$ detection sample." Note that the $z\sim2$ O$_{32}$ detection sample requires a robust detection of H$\alpha$, in addition to the lines needed to complete the O$_{32}$ vs. R$_{23}$ diagram, for accurate nebular dust corrections. Dust correcting the emission-lines is required for this diagram because the [O~\textsc{II}]$\lambda\lambda$3727,3730 and [O~\textsc{III}]$\lambda\lambda$4960,5008 lines are far apart in wavelength. This extra requirement does not remove any additional galaxies from the 181-galaxy sample. 

Our goal is to analyze a significantly more complete sample of MOSDEF star-forming galaxies. For this analysis, therefore, we exclude those galaxies identified as quiescent based on UVJ colors, as well as those flagged as AGNs. We also must limit our focus to galaxies with MOSFIRE spectroscopic redshifts, since we need a robust redshift for inclusion in spectral stacking.
If we exclude galaxies lacking a robust MOSFIRE redshift (referred to hereafter as $z_{\rm{MOSFIRE}}$), those flagged as AGN, and those identified as quiescent galaxies based on UVJ colors, there are 553 \textit{star-forming} galaxies in the $z\sim 2$ observed sample. 

To assemble a larger sample of $z\sim2$ MOSDEF star-forming galaxies, we relax the criteria for individual emission-line detections. However, because this larger sample will be used to create spectral stacks of emission lines in key diagnostic diagrams, we do require \textit{coverage} (if not detection) of all relevant spectral features. Since not all 553 galaxies (based on their redshifts, and locations in the MOSFIRE spectroscopic field of view) have the required spectral coverage, we remove an additional 75 galaxies that do not have coverage of all four lines (i.e., H$\beta$, [O~\textsc{III}]$\lambda$5008, H$\alpha$, and [N~\textsc{II}]$\lambda$6585). Therefore, we have 478 galaxies available for stacking on the [N~\textsc{II}] BPT diagram. 
For the [S~\textsc{II}] BPT diagram, we require coverage of [S~\textsc{II}]$\lambda\lambda$6717,6730 instead of [N~\textsc{II}]$\lambda$6585, and find that 
there are 472 $z\sim2$ star-forming galaxies with a robust $z_{\rm{MOSFIRE}}$ and coverage of H$\beta$, [O~\textsc{III}]$\lambda$5008, H$\alpha$, and [S~\textsc{II}]$\lambda\lambda$6717,6730 that are available for stacking. 
Finally, we require coverage of [O~\textsc{II}]$\lambda\lambda$3727,3730, H$\beta$, [O~\textsc{III}]$\lambda\lambda$4960,5008, and H$\alpha$ for analysis on the O$_{32}$ vs. R$_{23}$ diagram. While H$\alpha$ is not explicitly included in the line ratios plotted on the O$_{32}$ vs. R$_{23}$ diagram, its detection is required for dust correction, as described above. There are 406 galaxies that meet these criteria.

The stacked samples more than triple the number of individual galaxies shown on the [N~\textsc{II}] and [S~\textsc{II}] BPT diagrams in previous MOSDEF studies \citep[e.g.,][]{sha15, run21, run22}.
We hereafter refer to these samples as the MOSDEF ``$z\sim2$ N2 stacked sample'', MOSDEF ``$z\sim2$ S2 stacked sample'', and MOSDEF ``$z\sim2$ O$_{32}$ stacked sample''. 
Table \ref{tab:mosdef_samples} summarizes the various MOSDEF samples introduced in this section, providing both the name of each and number of galaxies each contains. 
The distributions of redshifts for the multiple MOSDEF samples samples are shown in Figure \ref{fig:redshift_hists_stacks}. 
The methodology for creating spectral stacks is discussed in Section \ref{subsec:spectral_stacking}. 

\begin{table}
    \centering
    \begin{tabular}{rr}
        \multicolumn{2}{c}{MOSDEF Samples} \\
        \hline\hline
        Sample Name & Number of Galaxies \\
        (1) & (2) \\
        \hline
    Parent Sample & 3780 \\
    Observed Sample & 786 \\
    Star-forming Observed Sample & 615 \\
    N2 Stacked Sample & 478 \\
    S2 Stacked Sample & 472 \\
    O$_{32}$ Stacked Sample & 406 \\
    Spectroscopic Sample & 250 \\
    N2 BPT Detection Sample & 143 \\
    S2 BPT Detection Sample & 156 \\
    O$_{32}$ Detection Sample & 181 \\
        \hline
    \end{tabular}
    \caption{
  Col. (1): Name of each MOSDEF sample. All samples are introduced in Section \ref{subsec:mosdef_sample_stack_paper} with the exception of the $z\sim2$ star-forming observed sample introduced in Section \ref{subsec:mosdef_targeted_completeness}.
  Col. (2): Number of galaxies in that sample.}
    \label{tab:mosdef_samples}
\end{table}

\subsection{SED Fitting and Derived Properties} \label{subsec:sed_fitting}

The fields targeted by MOSDEF  (AEGIS, COSMOS, GOODS-N, GOODS-S, and UDS) are covered by multi-wavelength photometry enabling robust SED fitting. All fields have \textit{Spitzer}/IRAC coverage; however, the specific set of ground-based and \textit{HST} bands from the near-UV to near-IR varies from field to field \citep{ske14}. 
When available, we use emission-line subtracted photometry for SED fitting, because strong lines (e.g. H$\beta$, [O~\textsc{III}]$\lambda\lambda$4960,5008, and H$\alpha$) can bias the shape of the SED fit redward of the Balmer break. The bias introduced by these emission-lines causes the modeling to favor older stellar population ages than if fitting the stellar continuum alone. The emission-line contribution to the photometry is estimated from the MOSFIRE spectra. 
Therefore, we use emission-line corrected photometry when there is an available MOSFIRE redshift. Otherwise, we use the raw photometry. 

We use the SED-fitting code FAST \citep{kri09} to  model the multi-wavelength photometry of MOSDEF galaxies, adopting the Flexible Stellar Population Synthesis (FSPS) library from \citet{con10} and assuming a \citet{cha03} stellar initial mass function (IMF) and delayed-$\tau$ star-formation histories where SFR(SED) $\propto$ $t \times e^{-t/\tau}$. Here, $t$ represents the time since the beginning of star formation and $\tau$ represents the characteristic star-formation timescale. 
Following previous studies \citep{du18, red18}, we adopt two combinations of dust attenuation curves and metallicity: one assumes a \citet{cal00} dust attenuation curve with the metallicity fixed to the defined solar value in the \citet{con10} library (0.019), and a second assumes a Small Magellanic Cloud (SMC) attenuation curve with the metallicity fixed to 28\% solar. We hereafter refer to these grids as the ``Calzetti+solar grid'' and the ``SMC+subsolar grid''. 
In the modeling, we fit for several key  galaxy properties, including $t$, $\tau$, the level of dust attenuation ($A_{\rm{V}}$), $M_{\ast}$, SFR(SED), and sSFR(SED) (i.e., SFR(SED)/$M_{\ast}$).

Following \citet{du18}, we use the Calzetti+solar grid for galaxies with log$_{10}(M_{\ast}/M_{\odot}) \geq 10.45$ and the SMC+subsolar grid for galaxies with log$_{10}(M_{\ast}/M_{\odot}) < 10.45$. The mass dependent manner in which we assign metallicity is consistent with the existence of a mass-metallicity relationship (MZR) (e.g., \citealt{tre04, erb06a, and13, ste14, san15}).

While not estimated directly from SED fitting, we can obtain rest-frame UVJ colors by passing the best-fit rest-frame FAST SED through U, V, and J filter bandpasses. We use the IRAF \citep{tod86, tod93} routine \textit{sbands} to obtain rest-frame UVJ colors. 
The combination of the U$-$V and V$-$J colors can distinguish between red quiescent galaxies and dusty star-forming galaxies by breaking the degeneracy for age and reddening \citep{wil09}. Quiescent galaxies are located in the upper left portion of the U$-$V vs. V$-$J diagram. 
As stated in the Section \ref{subsec:mosdef_sample_stack_paper}, we remove galaxies that fall in the quiescent region from this study.

\subsection{Emission-line Fitting} \label{subsec:emline_fits}

Emission-line measurements are obtained using a custom IDL code \citep[][]{red15}, which has been used in many previous MOSDEF studies \citep[e.g.,][]{sha15, sha19, sha22, san16, san18, san20, san21, run21, run22}. Prior to fitting MOSDEF uses an ``optimal'' \citep{fre19} 2D to 1D extraction method and the spectra are slit-loss corrected (see \citealt{kri15} for details). 
A single Gaussian profile is fit to each emission-line in the spectra, where the FWHM and centroid of each Gaussian are allowed to vary for each line. The lower and upper limits for the FWHM estimate are determined by the instrumental resolution estimated by sky lines and highest S/N line (most often H$\alpha$ or [O~\textsc{III}]$\lambda$5008). Once the FWHM for the highest S/N line is determined, the FWHM for all other emission lines cannot be 0.5 \AA\ larger than that. 
The continuum is fixed to the best-fit model from the FAST SED modeling. Additionally, the Balmer lines (H$\alpha$ and H$\beta$) are corrected for the underlying contribution of stellar Balmer absorption using the best-fit SED from FAST. 
As discussed in \citet{run22}, the code has recently been updated to more accurately estimate the magnitude of the stellar Balmer absorption contribution. 
If the best-fit Gaussian model is not a good fit to the spectrum, the code will choose the integrated bandpass flux as the preferred emission-line flux estimate. 
Uncertainties on the line fluxes are obtained by perturbing and refitting the spectrum 1000 times. The reported uncertainty is taken from 68th-percentile confidence interval of the distribution of perturbed flux measurements \citep{red15}.

\subsection{Spectral Stacking} \label{subsec:spectral_stacking}

Composite spectra were created using the methodology described in \citet{san18}. To summarize, each galaxy spectrum was shifted into the rest frame and converted from flux density to luminosity density using $z_{\rm{MOSFIRE}}$ and resampled onto a uniform wavelength grid. 
The spacing between each wavelength element is 0.5 \AA. In previous work \citep[e.g.,][]{san18,san21}, the luminosity of a strong emission line (e.g., [O~\textsc{III}]$\lambda$5008 or H$\alpha$) was used to normalize individual galaxy spectra. In this study, we normalized the spectra by SFR(SED) instead of H$\alpha$ emission-line luminosity because we included galaxies with S/N$_{\rm{H}\alpha} < 3$ and only required coverage (not detections) for the rest-optical emission lines we aimed to measure in stacked spectra. 
The individual spectra were median combined at each wavelength element. 
As a sanity check, we also tried normalizing the spectra by H$\alpha$ luminosity before stacking using the subset of galaxies with S/N$_{\rm{H\alpha}} \geq 3$. The emission-line ratios of the H$\alpha$ and SFR(SED) normalized stacks agreed within the uncertainties. This agreement demonstrates that normalizing by SFR(SED) is a suitable substitution for normalizing by H$\alpha$ luminosity when the latter has not been measured. Furthermore, SFR(SED) and SFR(H$\alpha$) have been shown to be correlated in previous work (e.g., \citealt{red15, shi16}), justifying this substitution.

For all analyses presented in this work, stacked spectra were constructed in 10 bins of $M_{\ast}$ with each bin containing approximately the same number of galaxies. 
Emission-line luminosities of stacked spectra were measured with the same methodologies used for fitting the individual spectra. Line flux uncertainties were estimated using a Monte Carlo technique where we perturbed the stellar masses of the individual galaxies within the uncertainties, divided the sample into the same $M_{\ast}$ bins based on the perturbed stellar masses, and refit the stacked spectra. This process was repeated 100 times, and the 68th-percentile confidence interval of fitted emission-line fluxes was used to estimate the line-flux uncertainty for each emission line. 
In each bin, the median Balmer absorption luminosity from the individual galaxies was used to correct the H$\alpha$ and H$\beta$ luminosities for the underlying stellar Balmer absorption. 

As stated above, for stacking on the [N~\textsc{II}] BPT diagram, we utilized all 478 $z\sim2$ star-forming galaxies (i.e., the N2 stacked sample) with a robust $z_{\rm{MOSFIRE}}$ and coverage of H$\beta$, [O~\textsc{III}]$\lambda$5008, H$\alpha$, and [N~\textsc{II}]$\lambda$6585. Of the 478 galaxies, 432 have S/N$_{\rm{H\alpha}} \geq 3$ with coverage of the remaining lines.
For the [S~\textsc{II}] BPT diagram there are 472 $z\sim2$ star-forming galaxies (i.e., the S2 stacked sample) with a robust $z_{\rm{MOSFIRE}}$ and coverage of H$\beta$, [O~\textsc{III}]$\lambda$5008, H$\alpha$, and [S~\textsc{II}]$\lambda\lambda$6718,6733 that are included in the stacking. 
Lastly, there are 406 galaxies (i.e., the O$_{32}$ stacked sample) with a robust $z_{\rm{MOSFIRE}}$ and coverage of [O~\textsc{II}]$\lambda\lambda$3727,3730, H$\beta$, [O~\textsc{III}]$\lambda\lambda$4960,5008, and H$\alpha$ required for dust-corrected emission-line ratios on the O$_{32}$ vs. R$_{23}$ diagram.

\begin{figure*}
    \includegraphics[width=0.98\linewidth]{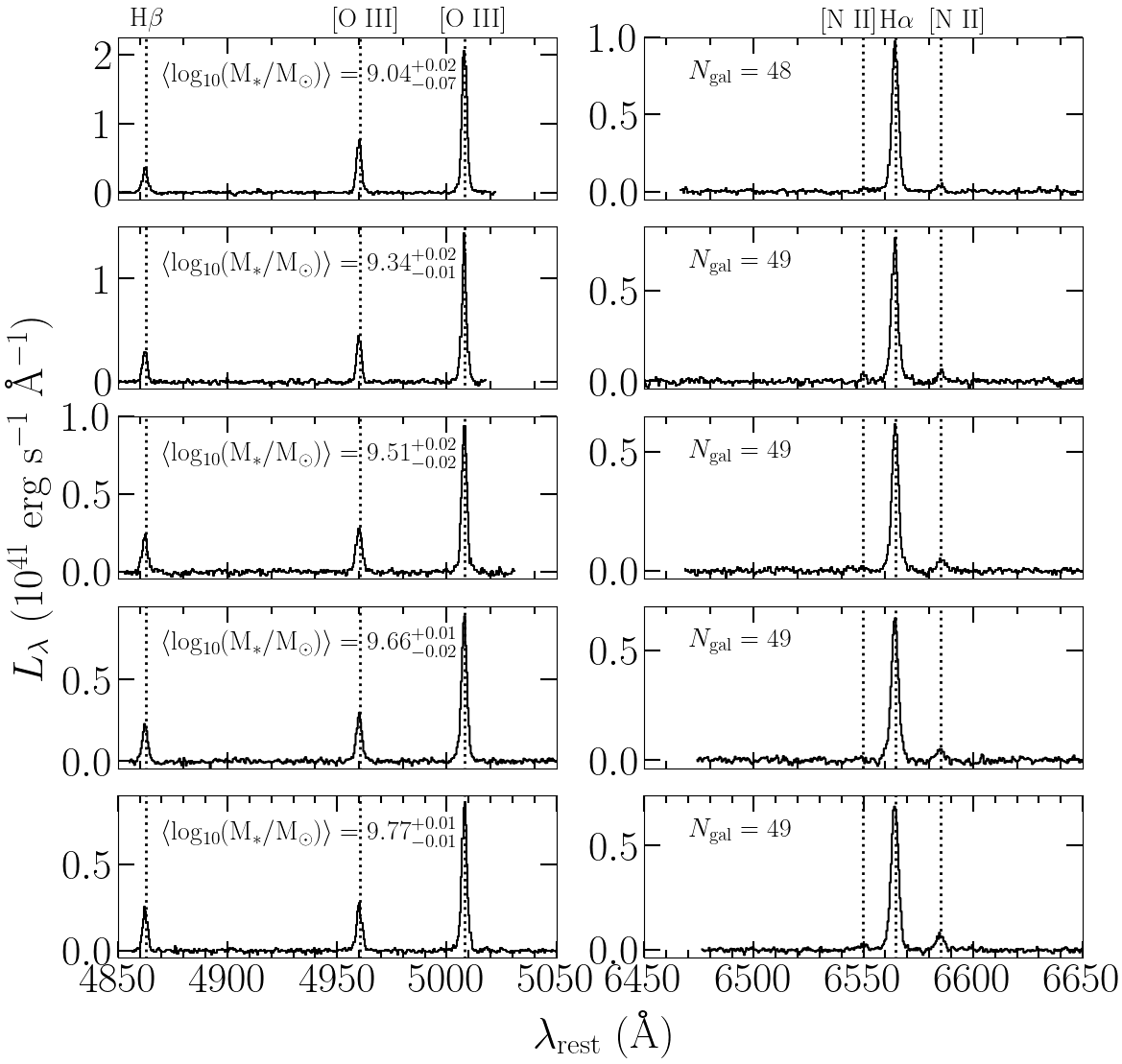}
    \caption{Composite spectra for the five lower-$M_{\ast}$ bins in the $z\sim2$ N2 stacked sample. The dotted vertical lines identify the rest-optical emission-lines in the stacks. The left panel shows H$\beta$ and [O~\textsc{III}]$\lambda\lambda$4960,5008 while the right panel shows H$\alpha$ and [N~\textsc{II}]$\lambda\lambda$6550,6585. The emission lines are labeled in the top panels. The average $M_{\ast}$ for each bin is given in the left panels while the number of galaxies in each bin ($N_{\rm{gal}}$) is given in the right panels. Note that the limits of the y-axis are adjusted in each panel to best show the emission lines. Also note that the emission lines show a general decrease in flux as the bins increase in $M_{\ast}$, most pronounced for H$\beta$ and [O~\textsc{III}]$\lambda\lambda$4960,5008.}
    \label{fig:composite_spectra_1}
\end{figure*}

\begin{figure*}
    \includegraphics[width=0.98\linewidth]{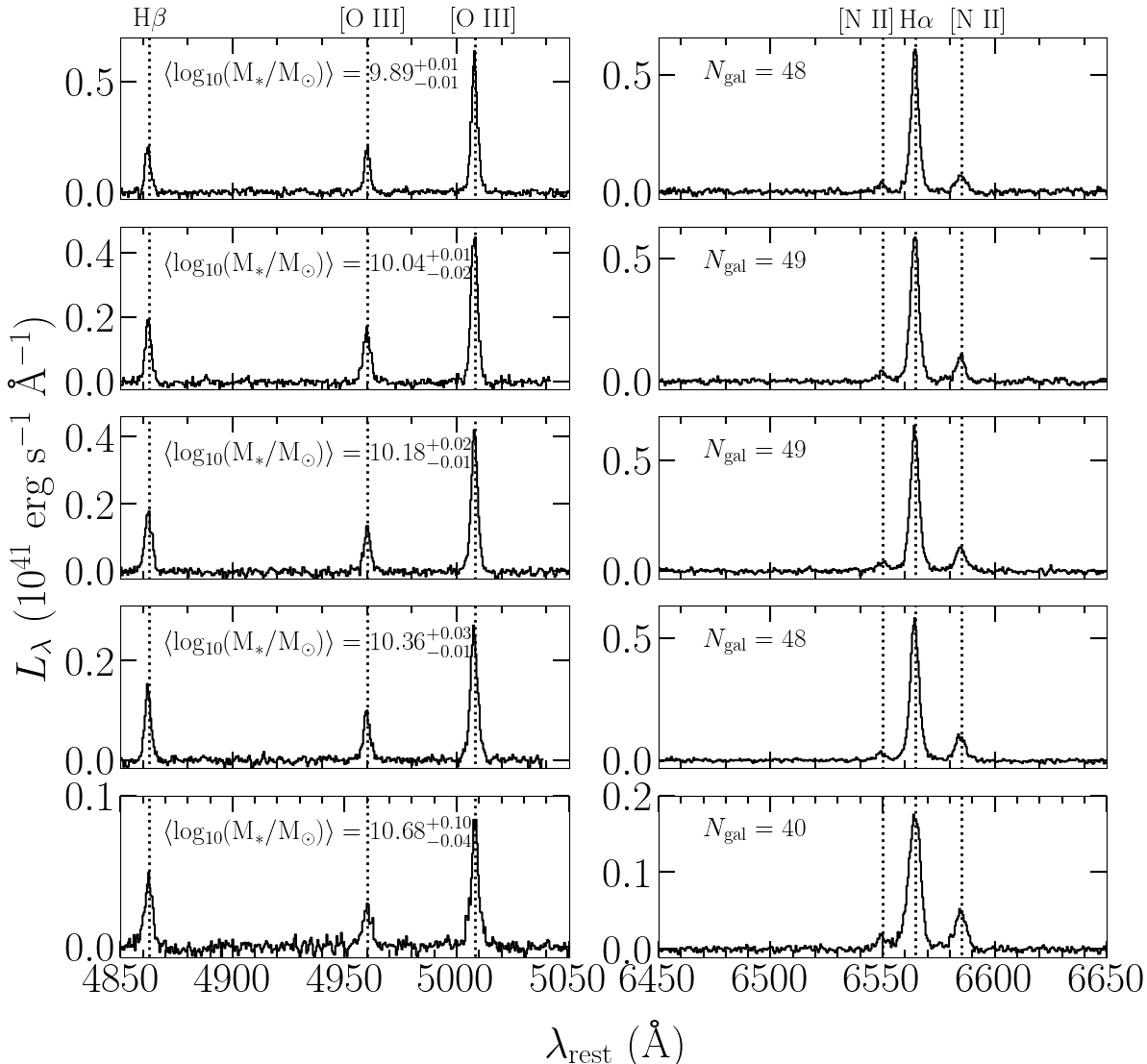}
    \caption{Same as Figure \ref{fig:composite_spectra_1} but displaying the composite spectra for the five higher-$M_{\ast}$ bins in the $z\sim2$ N2 stacked sample.}
    \label{fig:composite_spectra_2}
\end{figure*}

Figures \ref{fig:composite_spectra_1} and \ref{fig:composite_spectra_2} show the composite spectra for the 10 bins that comprise the N2 stacked sample (five mass bins in each figure). In both figures, the left panel shows H$\beta$ and [O~\textsc{III}]$\lambda\lambda$4960,5008 (observed $H$-band)  while the right panel shows H$\alpha$ and [N~\textsc{II}]$\lambda\lambda$6550,6585 (observed $K_{\rm{s}}$-band). The number of galaxies and the mean $M_{\ast}$ in each bin are also indicated. The axis limits in a given band are the same for all panels (i.e., mass bins) in Figures \ref{fig:composite_spectra_1} and \ref{fig:composite_spectra_2}, to enable a visual comparison of the line flux strengths across mass bins.

\begin{figure*}
    \includegraphics[width=0.49\linewidth]{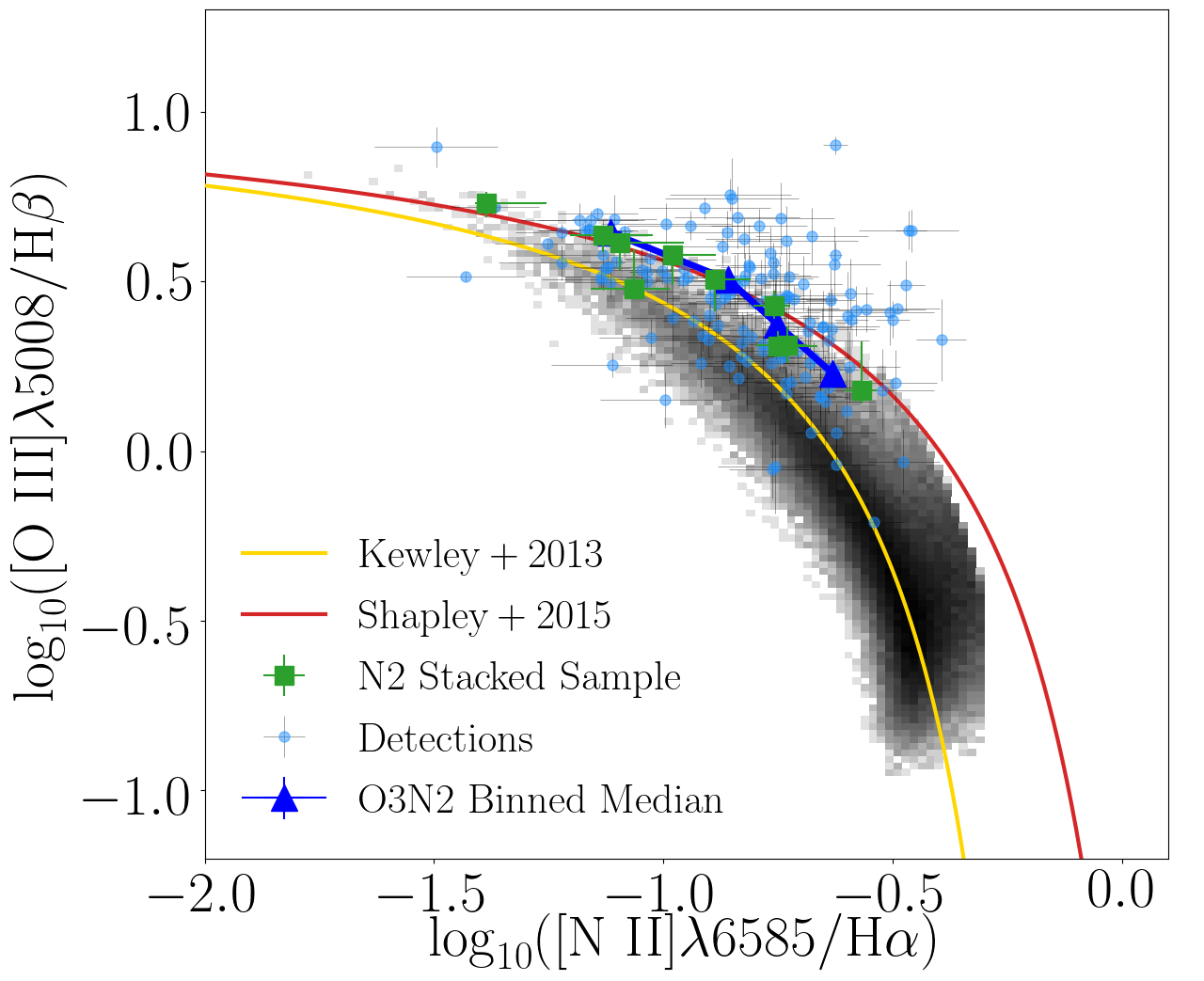}
    \includegraphics[width=0.49\linewidth]{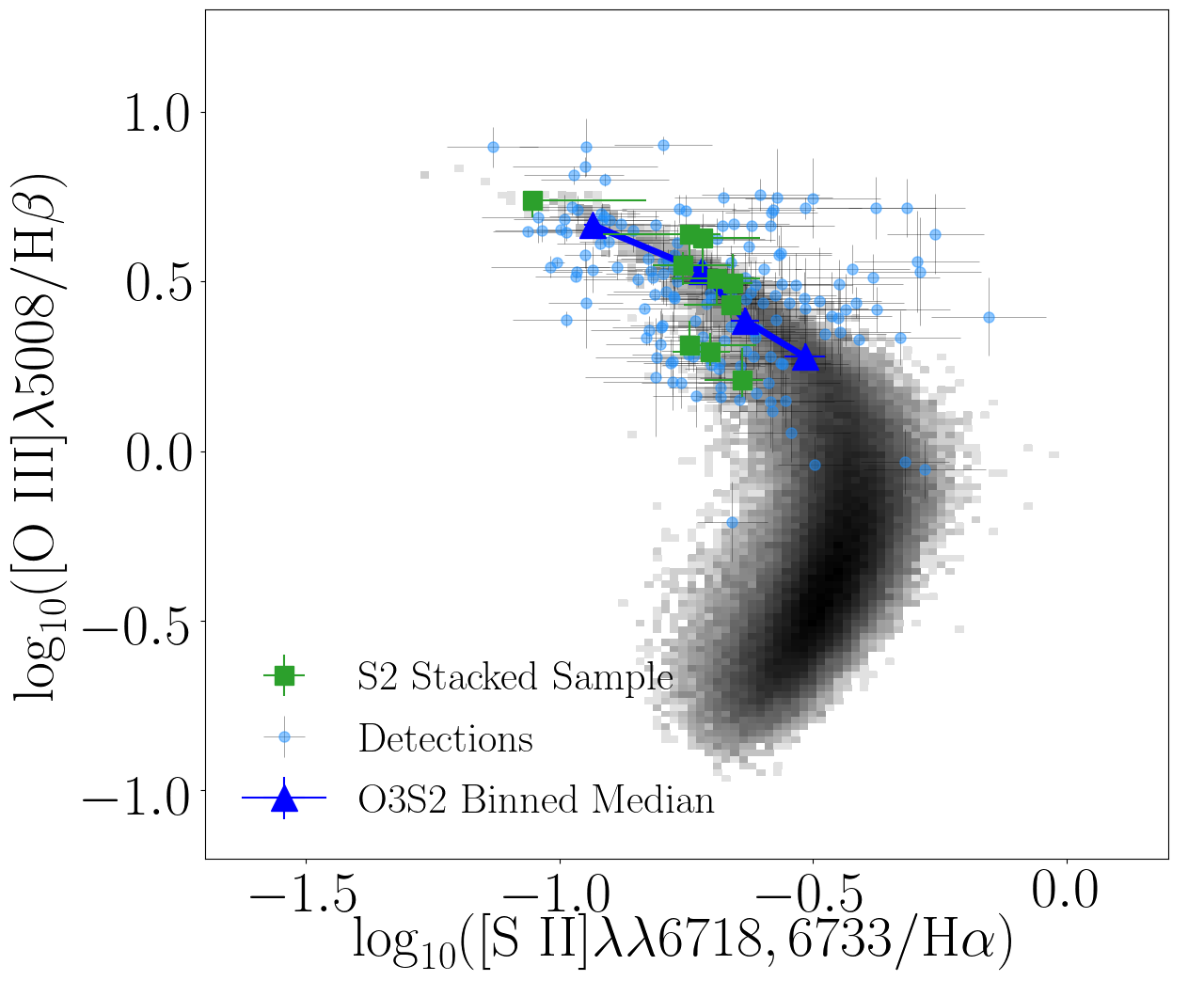}
    \includegraphics[width=0.49\linewidth]{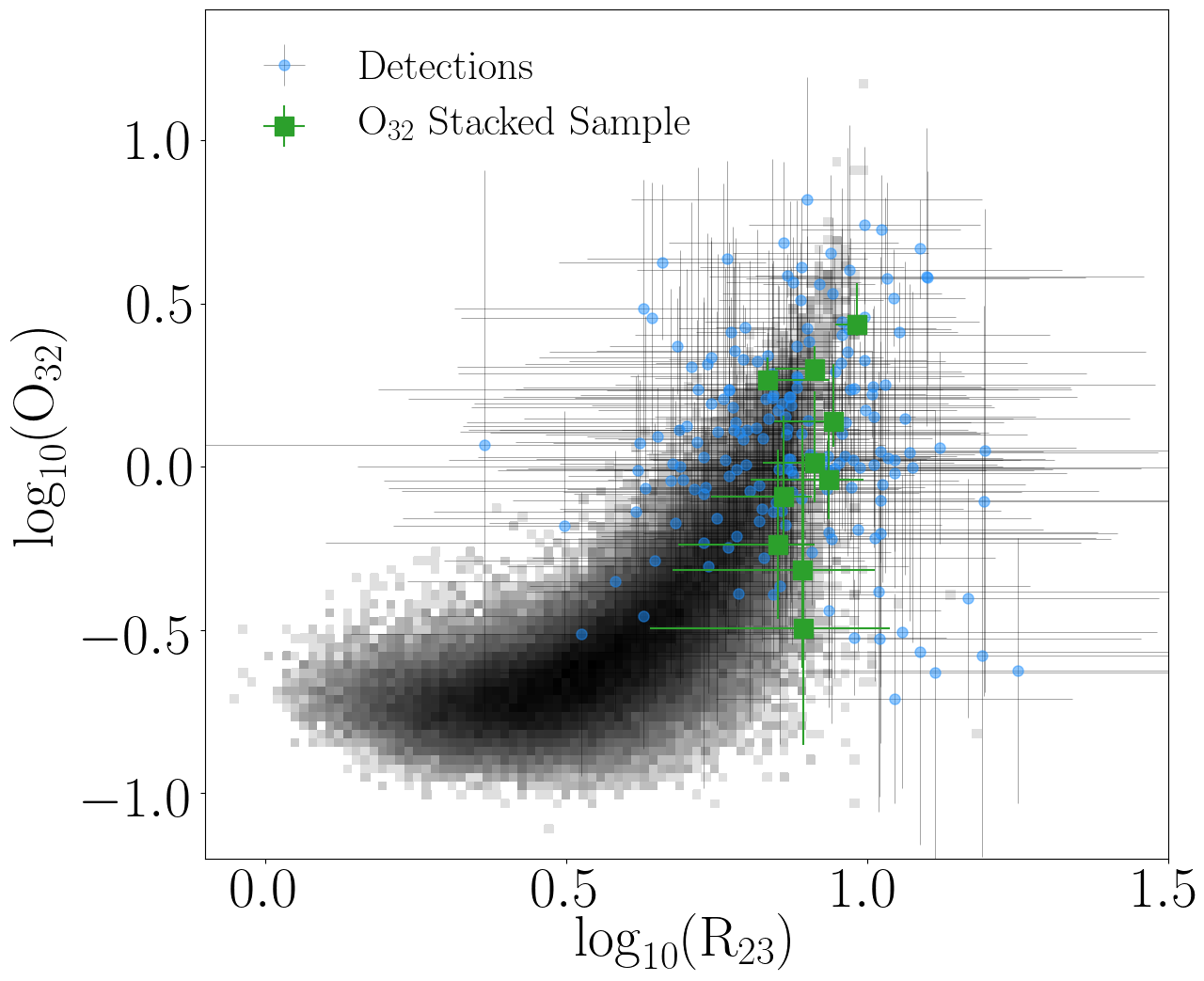}
    \caption{The [N~\textsc{II}] BPT (top left), [S~\textsc{II}] BPT (top right), and O$_{32}$ vs. R$_{23}$ (bottom) diagrams. 
    The N2 and S2 $z\sim2$ stacked samples are shown with green squares. 
    The grey 2D histograms show the local SDSS $z\sim0$ star-forming galaxies. 
    The blue points identify the 143, 156, and 181 galaxies with 3$\sigma$ S/N for all relevant lines on the [N~\textsc{II}] BPT, [S~\textsc{II}] BPT, and O$_{32}$ vs. R$_{23}$ diagrams, respectively. An additional requirement for a robust H$\alpha$ detection is needed for dust corrections on the O$_{32}$ vs. R$_{23}$ diagram. The blue triangles are the binned medians from \citet{run22}, binned in O3N2 (O3S2) on the [N~\textsc{II}] ([S~\textsc{II}]) BPT diagram. 
    The red curve on the [N~\textsc{II}] BPT diagram is a fit to early MOSDEF data from \citet{sha15}, while the yellow curve is a fit to the SDSS $z\sim0$ star-forming locus \citep{kew13}.}
    \label{fig:bpt_plots}
\end{figure*}

\begin{figure*}
    \centering
     \subfloat[]{
       \includegraphics[width=0.32\linewidth]{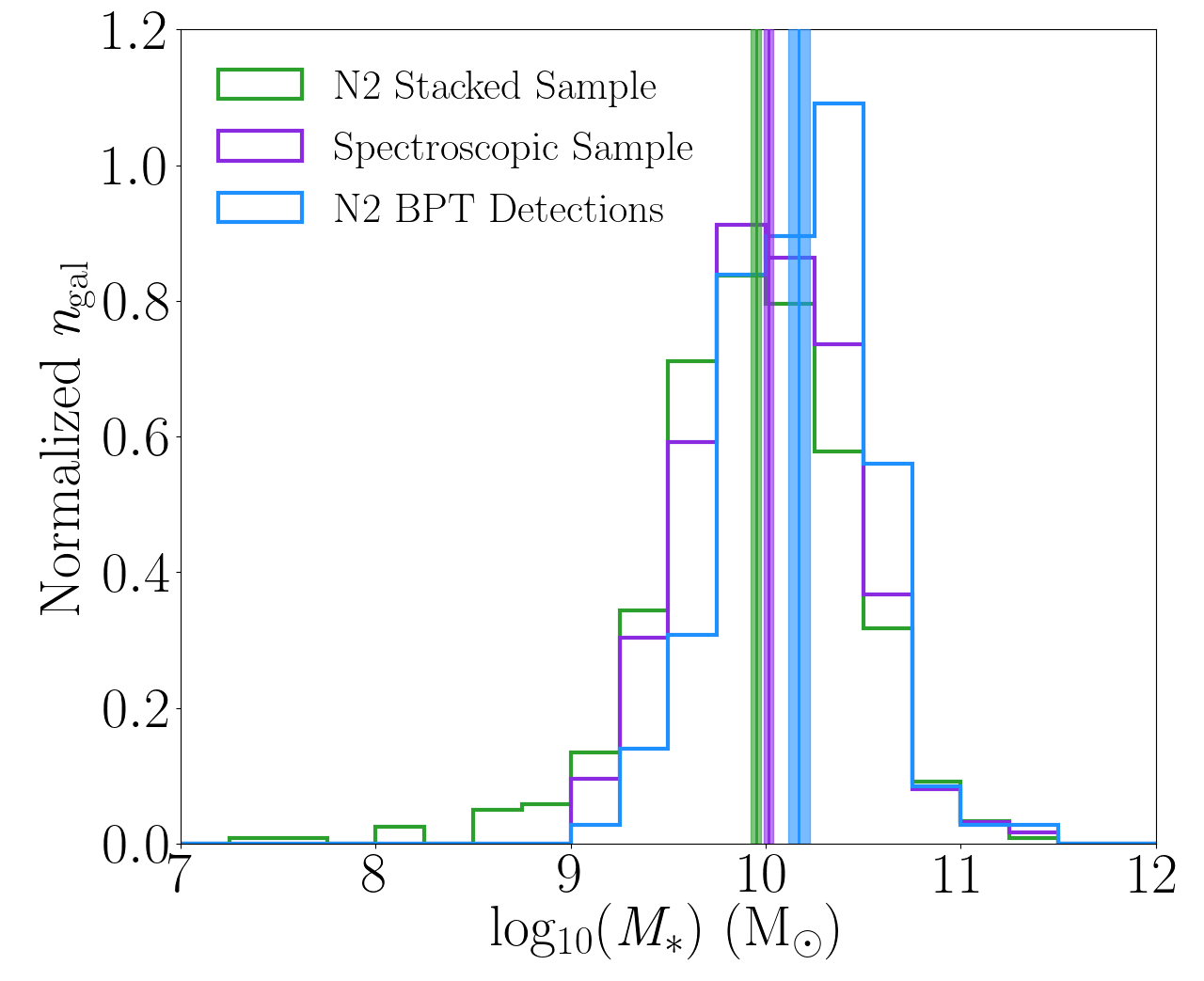}
     }
     \hfill
     \subfloat[]{
       \includegraphics[width=0.32\linewidth]{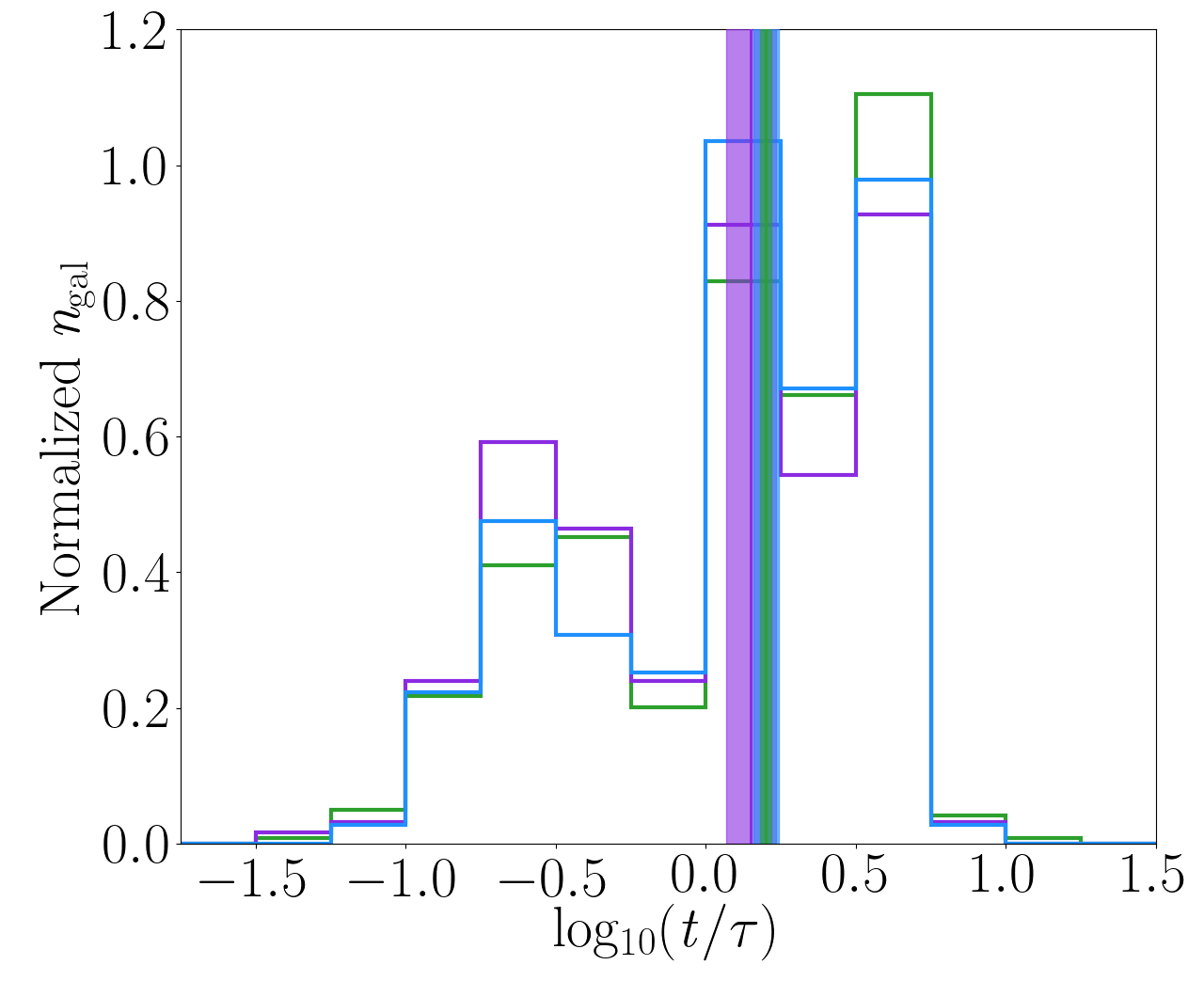}
     }
     \hfill
     \subfloat[]{
       \includegraphics[width=0.32\linewidth]{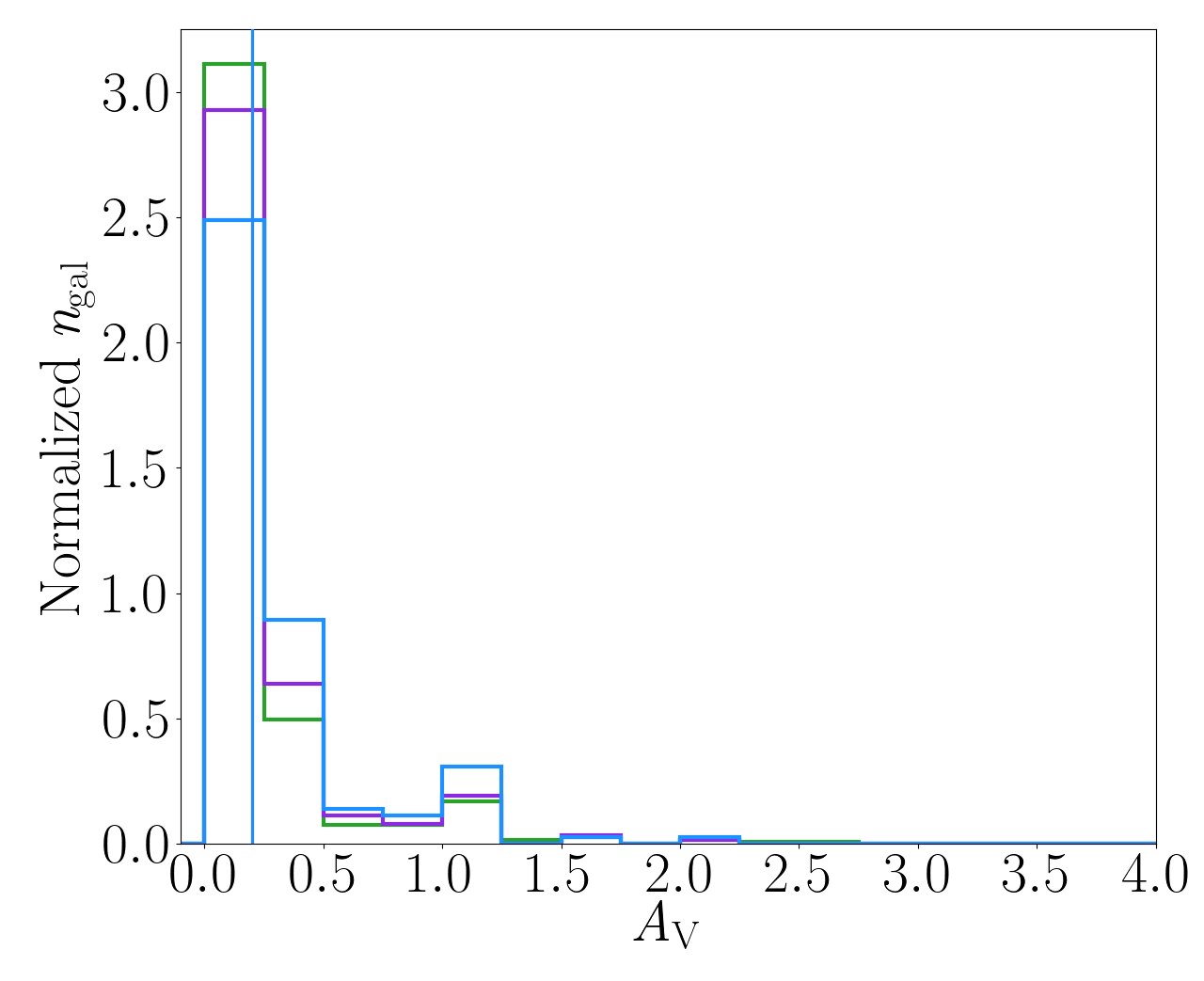}
     }
     \hfill
     \subfloat[]{
       \includegraphics[width=0.32\linewidth]{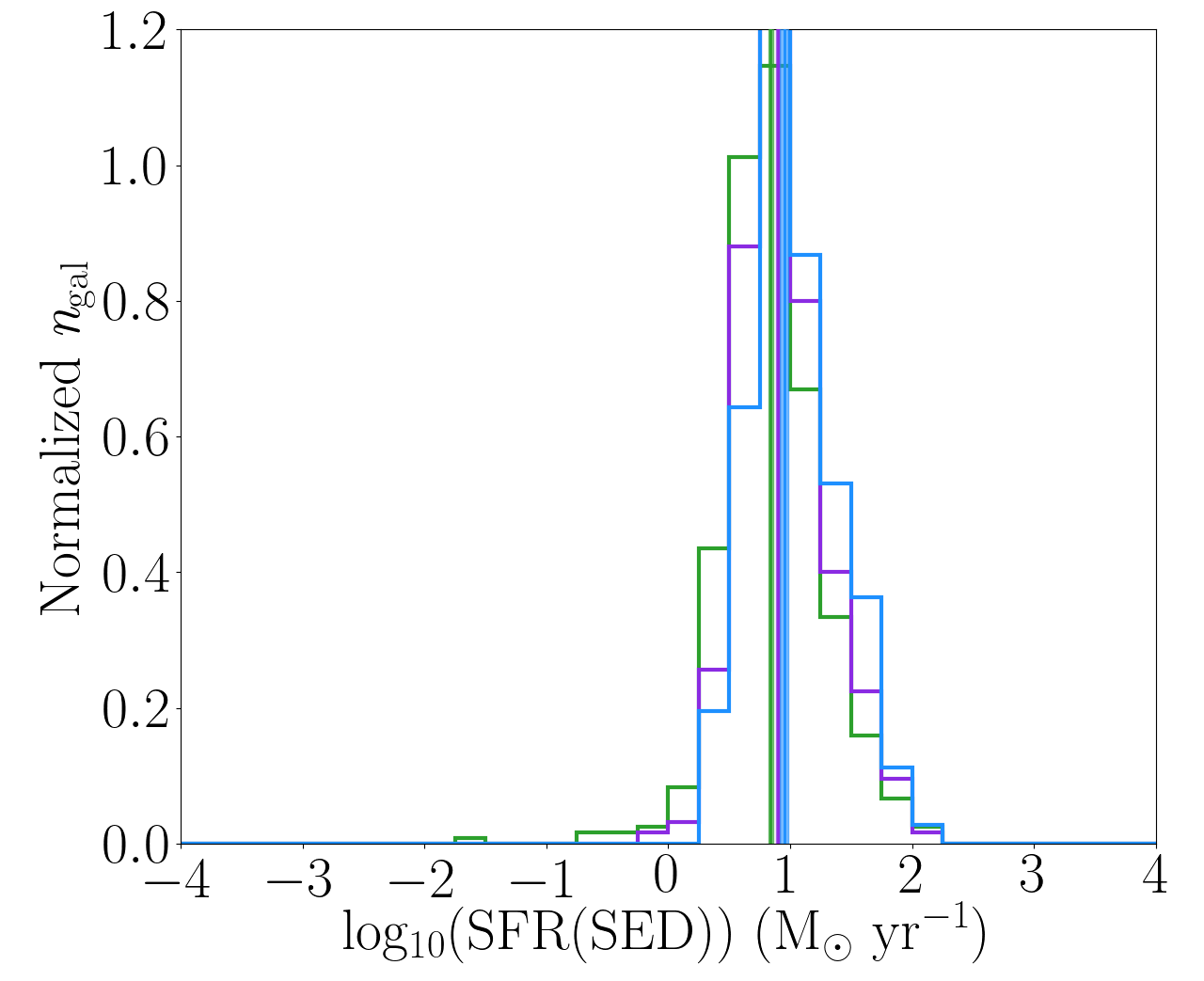}
     }
     \hfill
     \subfloat[]{
       \includegraphics[width=0.32\linewidth]{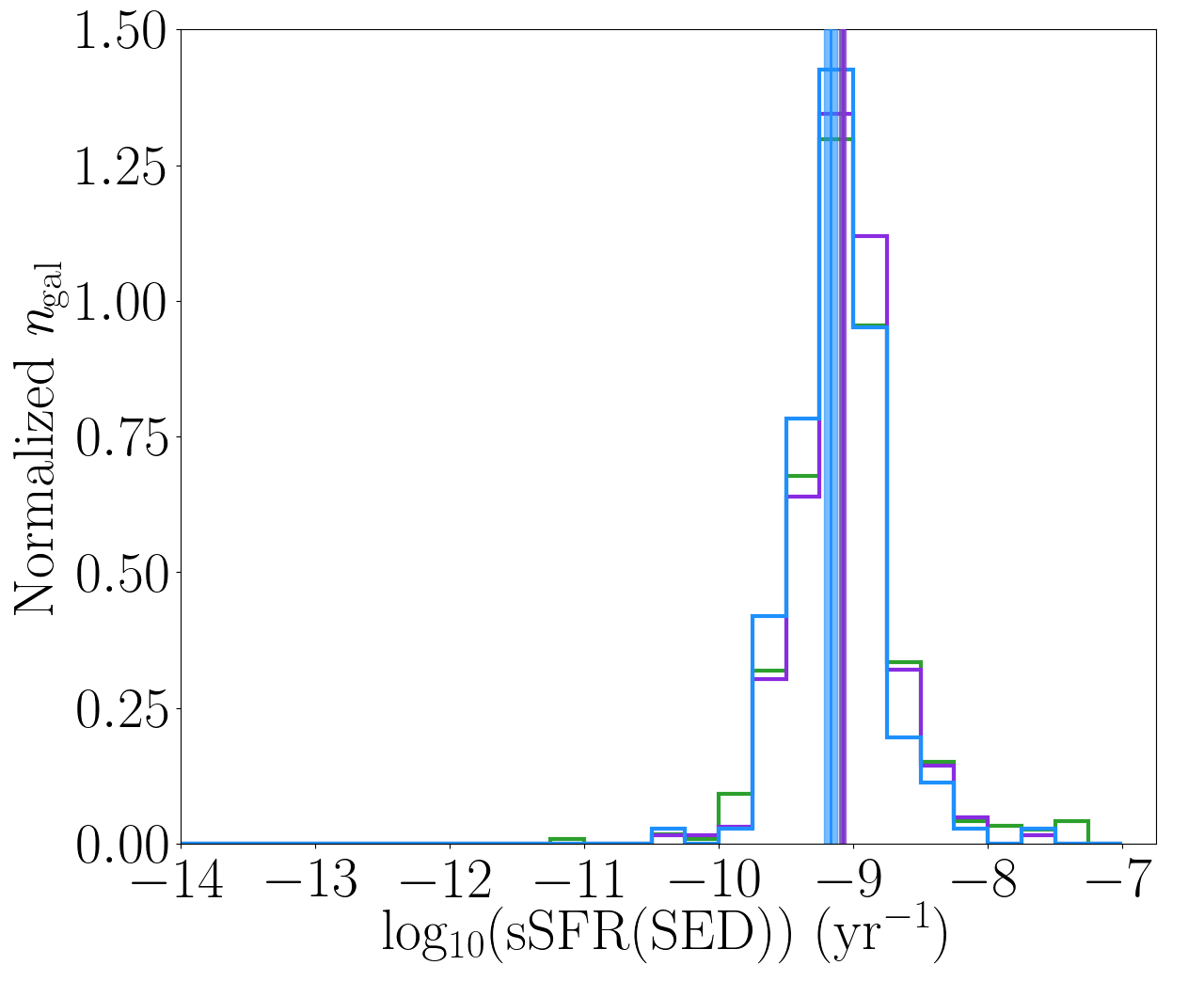}
     }
     \hfill
     \subfloat[]{
       \includegraphics[width=0.32\linewidth]{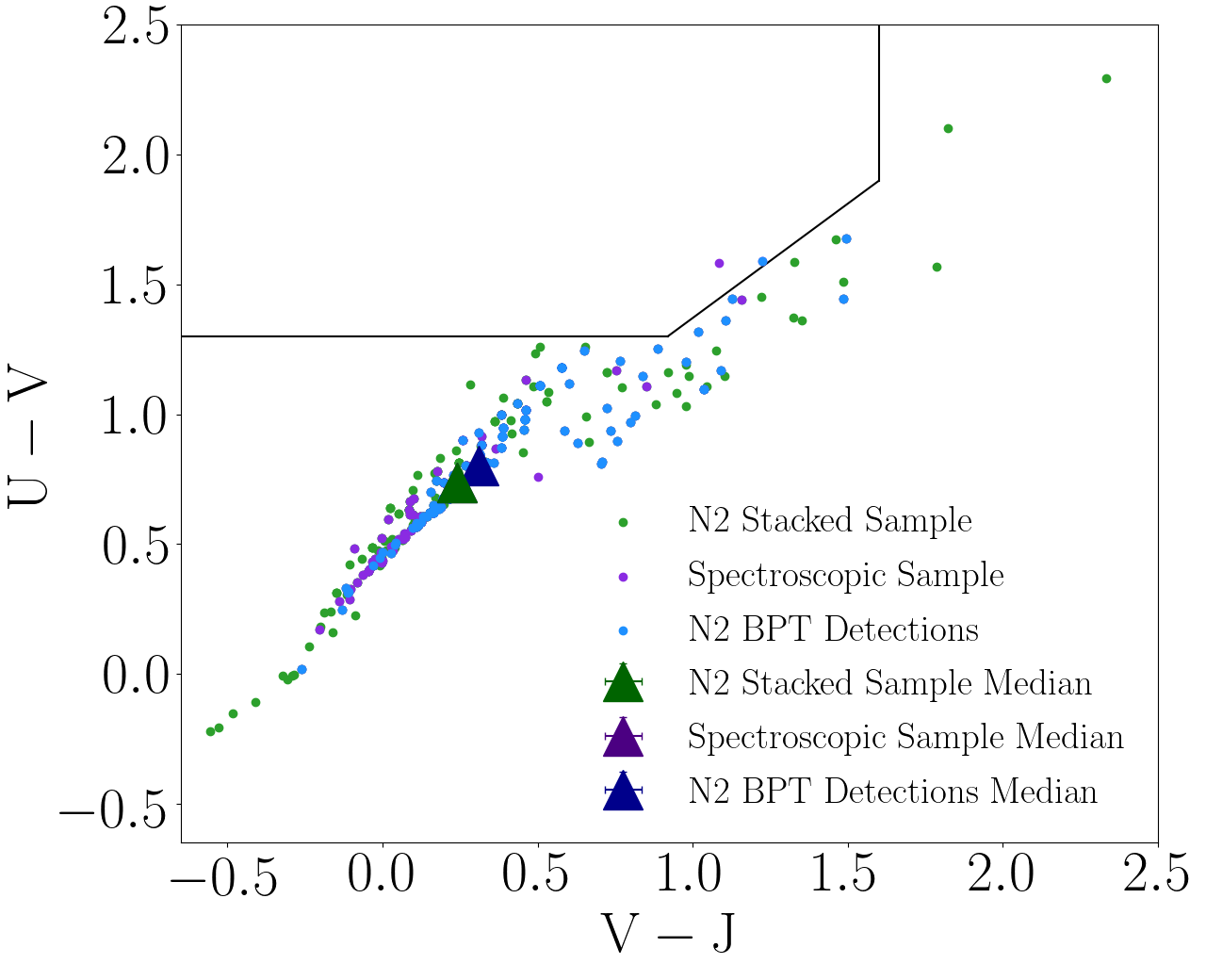}
     }
    \caption{Distribution of physical properties and sample medians with 1$\sigma$ uncertainties for the MOSDEF N2 $z\sim2$ stacked sample (green), the $z\sim2$ spectroscopic sample (purple), and the 143 galaxy subset of the $z\sim2$ spectroscopic sample shown on the [N~\textsc{II}] BPT diagram in Figure~\ref{fig:bpt_plots} (blue), i.e., the N2 BPT detection sample. The values for the sample medians are given in Table \ref{tab:z2_stack_zspec_sample_properties}. 
    For the panels with 1D histograms, the y-axis is normalized so the area under the histogram for each sample adds up to one. 
    Distributions for the following galaxy properties are shown: (a) $M_{\ast}$, (b) log$_{10}$($t/\tau$) of the stellar population assuming a delayed-$\tau$ star formation model, (c) $A_{\rm{V}}$, (d) SFR(SED), (e) sSFR(SED), (f) the UVJ diagram. The box on the UVJ diagram separates the quiescent region (upper left) from the star-forming region (bottom half and upper right). Additionally, the sample medians on the UVJ diagram are dark green (N2 $z\sim2$ stacked sample), dark purple ($z\sim2$ spectroscopic sample), and dark blue (N2 BPT detection sample) for easier visibility. Note that the N2 $z\sim2$ stacked sample and $z\sim2$ spectroscopic sample have the same U$-$V and V$-$J median values. The galaxy properties shown here are estimated using emission-line corrected photometry for all galaxies in the figure. We note that the sample medians in panel (c) overlap.}
    \label{fig:z2_stack_spec_sample_properties}
\end{figure*}

\subsection{SDSS Comparison Sample} \label{subsec:local_sdss_sample_stacking}

We compare the emission lines of our various $z\sim2$ MOSDEF samples on the [N~\textsc{II}] and [S~\textsc{II}] BPT diagrams with similar measurements from galaxies in the local universe. For these comparisons, we utilized archival emission-line measurements from the SDSS Data Release 7 (DR7; \citealt{aba09}), specifically from the MPA-JHU DR7 release of spectrum measurements\footnote{https://wwwmpa.mpa-garching.mpg.de/SDSS/DR7/}. 
We restricted the SDSS sample to a redshift range of $0.04 \leq z \leq 0.10$ and removed galaxies with a poor $M_{\ast}$ estimate. Additionally, AGN were removed using equation 1 from \citet{kau03} or if N2 $>$ 0.5. We employed a S/N cut of 3 to the relevant emission-lines on the [N~\textsc{II}], [S~\textsc{II}] BPT, and O$_{32}$ vs. R$_{23}$ diagrams. 
Similar to the MOSDEF sample, we required S/N$_{\rm{H}\alpha}\geq3$ for the sample shown on the O$_{32}$ vs. R$_{23}$ diagram to perform the dust corrections. 
These criteria resulted in SDSS samples of 96,346, 95,132, and 76,810 star-forming galaxies on the [N~\textsc{II}] BPT, [S~\textsc{II}] BPT, and O$_{32}$ vs. R$_{23}$ diagrams, respectively.

\section{Results} \label{sec:results}

In this section, we compare the multiple different MOSDEF samples introduced in Section \ref{subsec:mosdef_sample_stack_paper}. In Section \ref{subsec:stack_emlines}, we analyze the emission-line properties of the MOSDEF $z\sim2$ stacked and spectroscopic samples on the [N~\textsc{II}] BPT, [S~\textsc{II}] BPT, and O$_{32}$ vs. R$_{23}$ diagrams. 
In Section \ref{subsec:stack_properties}, we compare the host galaxy properties of the MOSDEF $z\sim2$ stacked and spectroscopic samples using emission-line corrected photometry in the SED modeling. Additionally, we compare these samples with the MOSDEF $z\sim2$ samples from \citet{san21} and \citet{sha22} on diagrams correlating emission-line ratios with $M_{\ast}$.
Finally, Section \ref{subsec:parent_vs_targeted_samples} provides a comparison of the host galaxy properties of the MOSDEF $z\sim2$ parent and observed samples using emission-line uncorrected photometry in the SED modeling.

\subsection{Stacking on the BPT diagrams} \label{subsec:stack_emlines}

We show the $z\sim2$ N2, S2, and O$_{32}$ stacked samples in bins of $M_{\ast}$ on the [N~\textsc{II}] BPT (top left), [S~\textsc{II}] BPT (top right), and O$_{32}$ vs. R$_{23}$ (bottom) diagrams in Figure \ref{fig:bpt_plots}. Also included are the $z\sim2$ N2 BPT and S2 BPT detection samples from \citet{run22}
with galaxies binned according to log$_{10}$(O3N2) on the [N~\textsc{II}] BPT diagram and log$_{10}$(O3S2) on the [S~\textsc{II}] BPT diagram.
The $z\sim2$ O$_{32}$ detection sample is included on the O$_{32}$ vs. R$_{23}$ diagram. 
On the [N~\textsc{II}] BPT diagram, the sequence of $M_{\ast}$ bins has a similar distribution relative to the local sequence as observed in previous MOSDEF studies (i.e., Equation 1 from \citealt{sha15} fitting early MOSDEF data and the O3N2 binned median from \citealt{run22}). 
The bin in the $z\sim2$ N2 stacked sample with the lowest $M_{\ast}$ is located at the upper left (i.e., high O3 and low N2) of the diagram. This bin extends past the region covered by the O3N2 binned median from \citep{run22}. As bins in the $z\sim2$ N2 stacked sample increase in $M_{\ast}$, the sequence moves to lower O3 and higher N2. This trend is expected as position on the [N~\textsc{II}] BPT diagram is shown to correlate with $M_{\ast}$ (e.g., \citealt{mas16, run21}). 
For the N2 $z\sim2$ stacked sample, the perpendicular offset from the \citet{kew13} fit to the local SDSS sequence is 0.10 $\pm$ 0.04 dex. This distance is slightly lower, but agrees within the uncertainties to the smaller $z\sim2$ N2 BPT detection sample (0.12 $\pm$ 0.02). 

On the [S~\textsc{II}] BPT diagram, the $z\sim2$ S2 stacked sample has a similar shape to the O3S2 binned median for the $z\sim2$ S2 BPT detection sample. However, at high S2, the stacks are offset to lower O3 compared to the O3S2 sequence. Similar to what is observed in the [N~\textsc{II}] BPT diagram, the bin in the $z\sim2$ S2 stacked sample with the lowest $M_{\ast}$ is located at the upper left part of the diagram (i.e., high O3 and low S2). 
Both the $z\sim2$ S2 stacked and S2 BPT detection samples occupy the region of the local SDSS star-forming sequence. 
However, the lowest-mass bin of the $z\sim2$ S2 stacked sample does not extend past the $z\sim2$ S2 BPT detection sample sequence of O3S2 bins to higher O3 and lower S2 (unlike what is found on the [N~\textsc{II}] BPT diagram).

The $z\sim2$ O$_{32}$ stacked sample occupies a similar region of O$_{32}$ vs. R$_{23}$ space as the $z\sim2$ O$_{32}$ detection sample. Both samples are offset to higher R$_{23}$ compared to the local SDSS star-forming sequence. 
The photoionization models from \citet{run21} suggest that this offset is caused by the $z\sim2$ star forming galaxies having a lower stellar metallicity (i.e., a harder ionizing spectra) at fixed gas-phase oxygen abundance than galaxies at $z\sim0$. Many studies support the idea that $z\sim2$ star-forming galaxies contain a harder ionizing spectra at fixed gas-phase oxygen abundance \citep{ste14, ste16, str17,sha19, top20a, red21}. 

Figure \ref{fig:bpt_plots} shows that the median emission-line properties of the $z\sim2$ N2 BPT, S2 BPT, and O$_{32}$ detection samples are representative of the emission-line properties for the full stacked samples of galaxies with coverage of the emission-lines needed for each diagram. These results indicate that previous MOSDEF studies (e.g., \citealt{sha15, sha19, sha22, san16, san18, san20, san21, run21, top20a, top20b}) utilizing high S/N subsets of the $z\sim2$ MOSDEF survey most likely show little bias in the emission-line properties compared to the full sample.  
We discuss this topic more in Section \ref{sec:discussion}.

\begin{figure*}
    \includegraphics[width=0.49\linewidth]{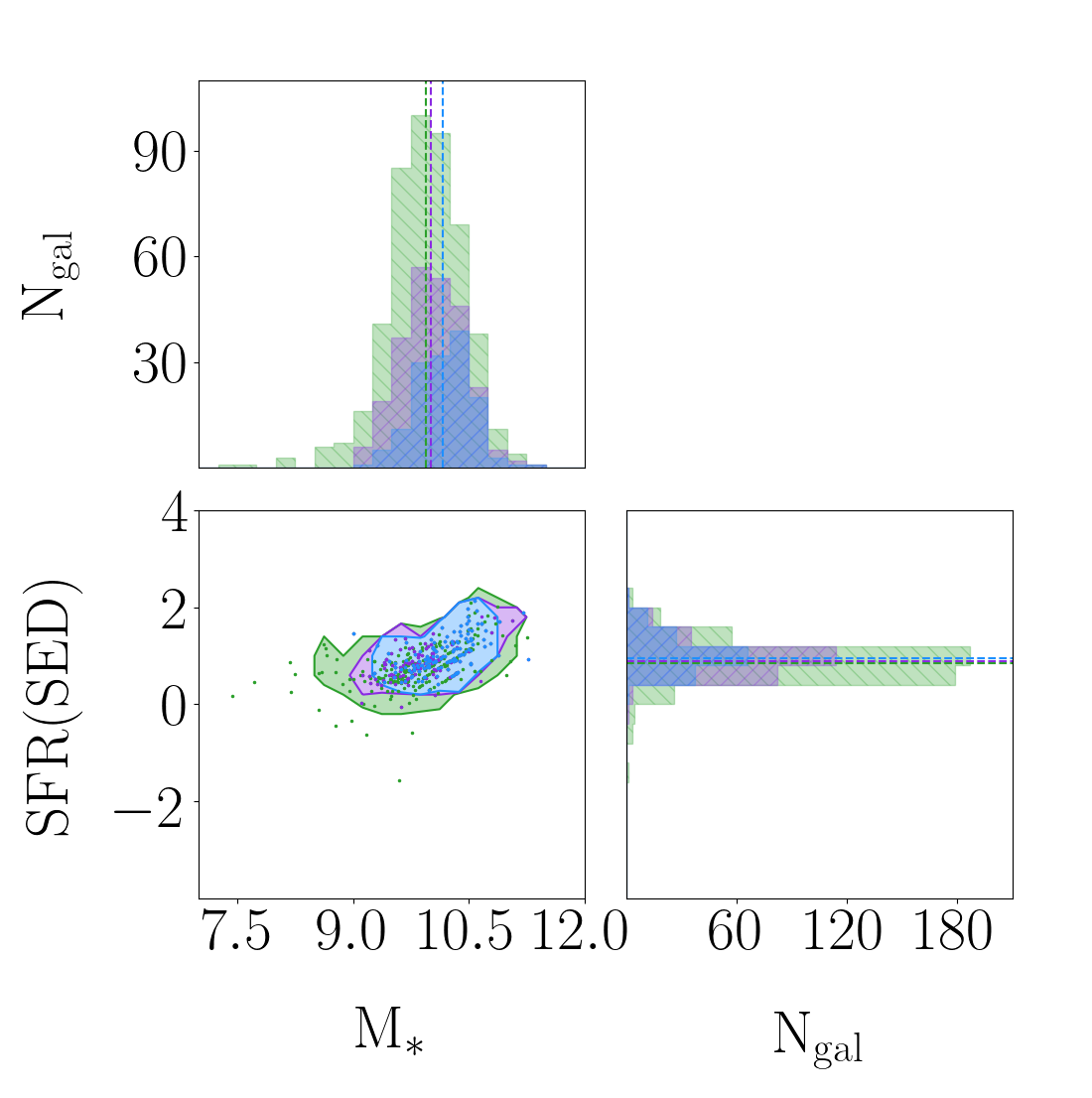}
    \includegraphics[width=0.49\linewidth]{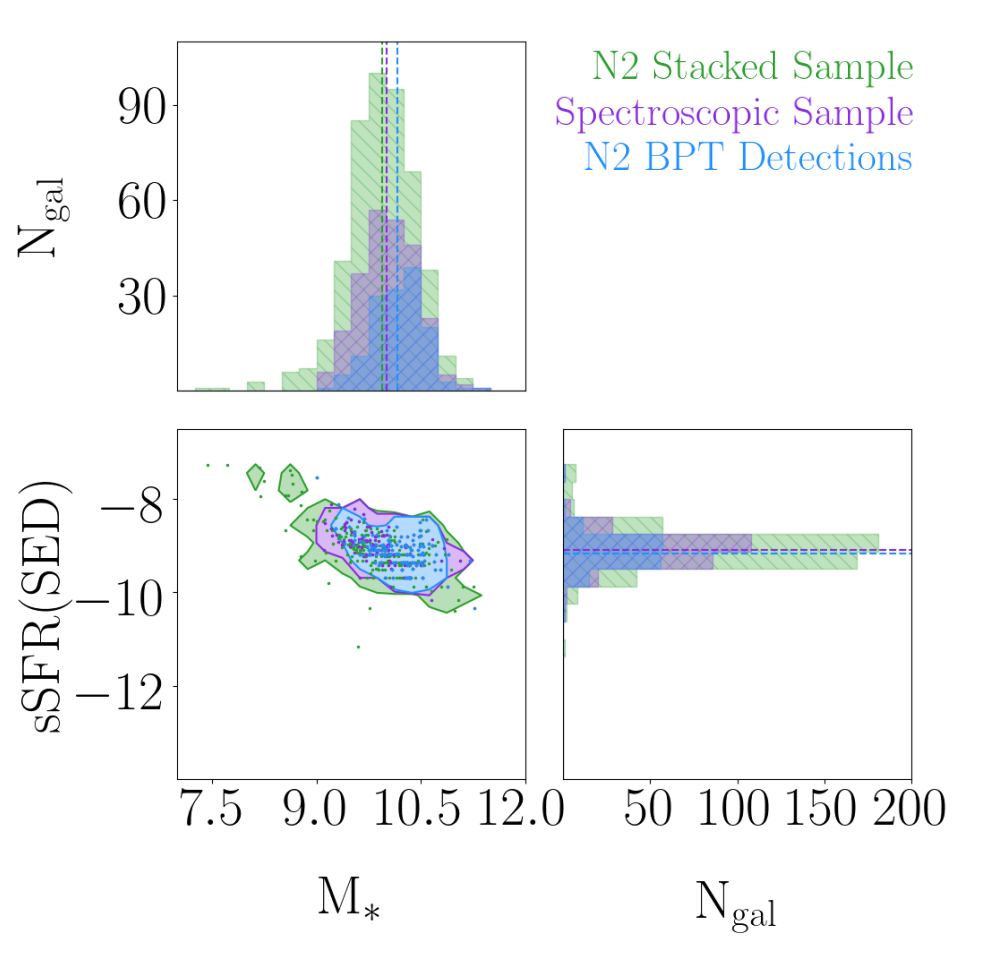}
    \includegraphics[width=0.49\linewidth]{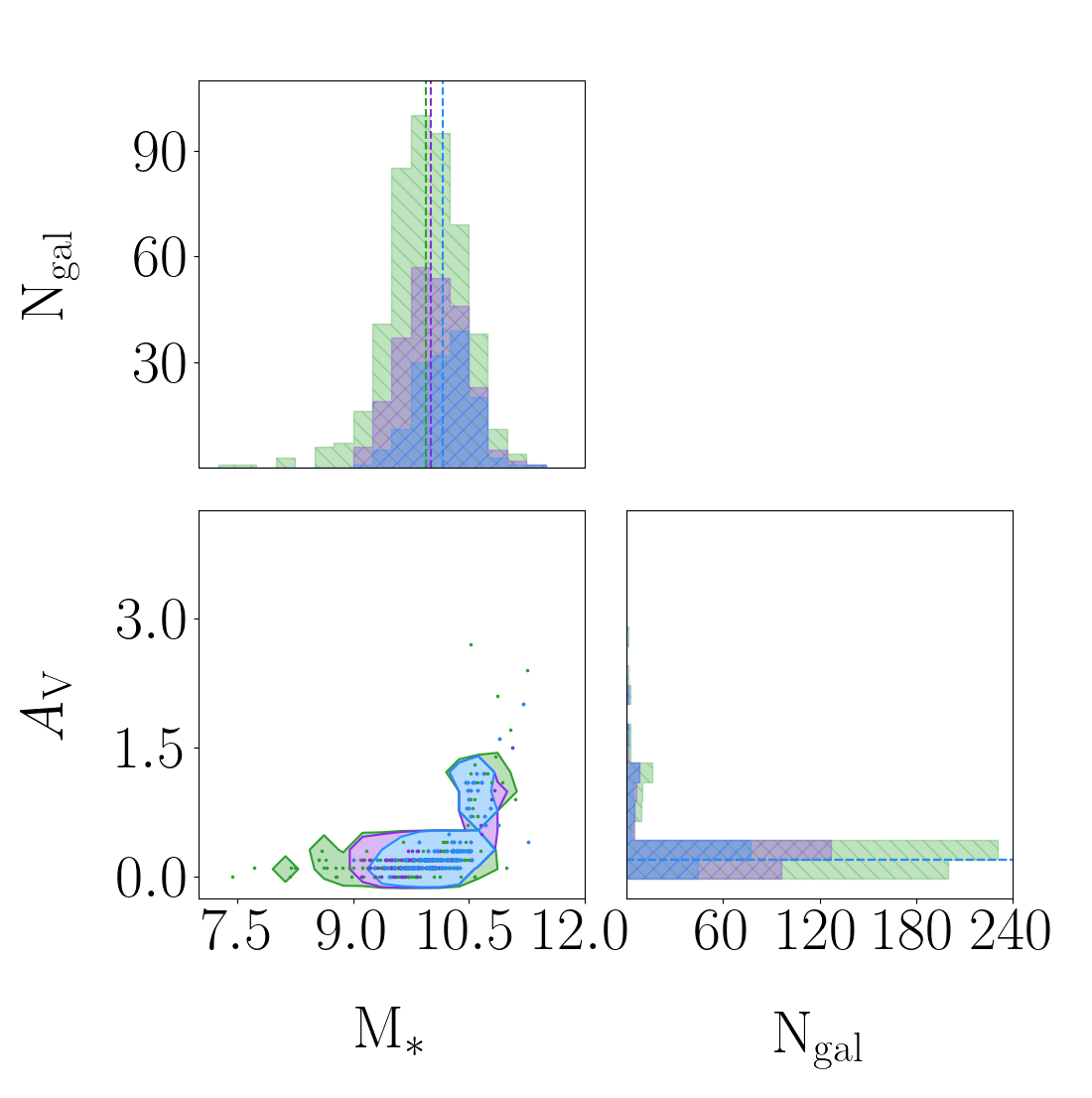}
    \includegraphics[width=0.49\linewidth]{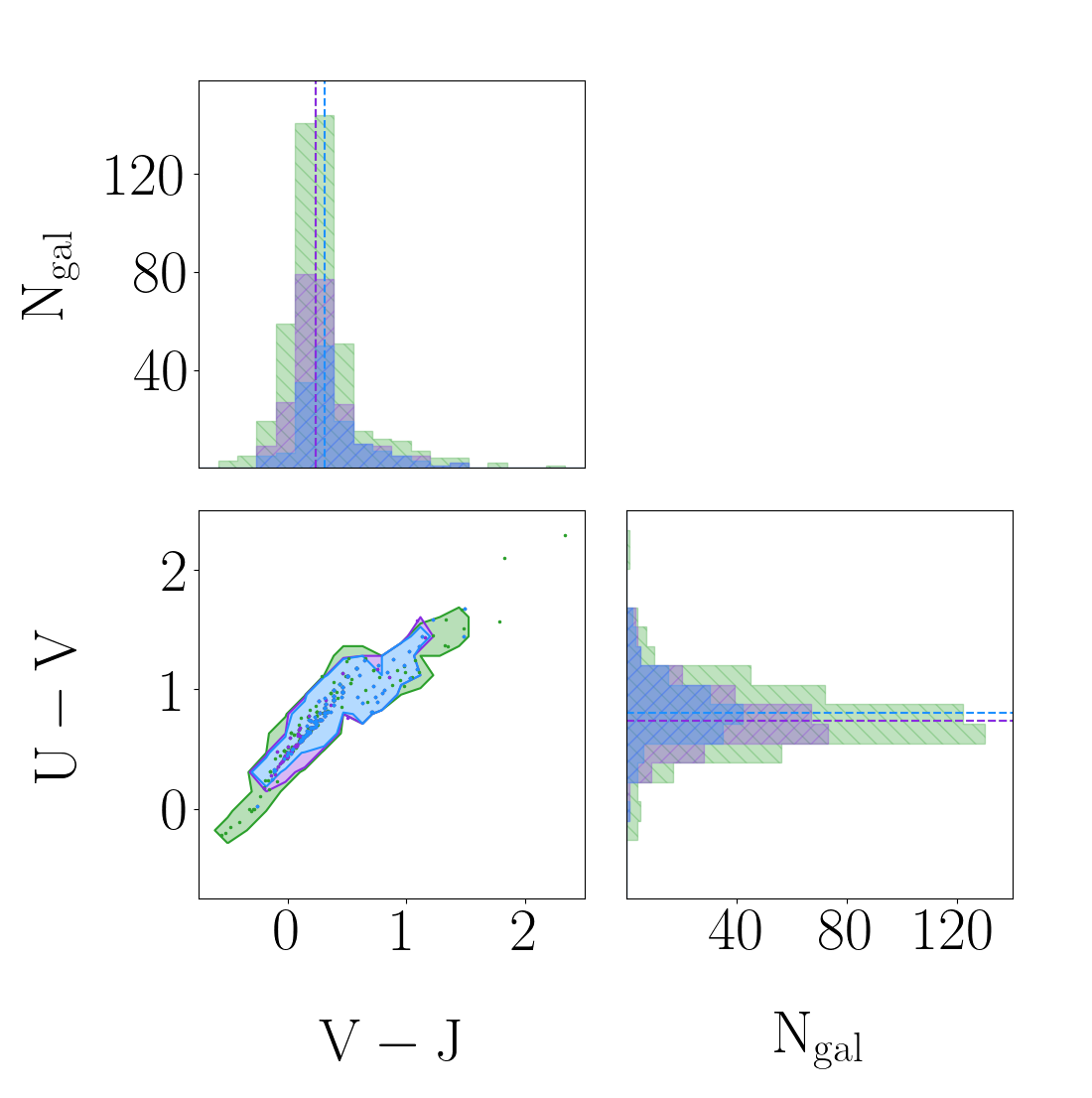}
    \caption{Corner plots for SFR(SED) vs. $M_{\ast}$ (upper left), sSFR(SED) vs. $M_{\ast}$ (upper right), $A_{\rm{V}}$ vs. $M_{\ast}$ (bottom left), and U$-$V vs. V$-$J (bottom right). Shown on these diagrams are the $z\sim2$ stacked sample (green), $z\sim2$ spectroscopic sample (purple), and $z\sim2$ N2 BPT detections (blue). The galaxy properties shown here are estimated using emission-line corrected photometry for all galaxies in the figure. The parameter distributions and sample medians are shown in the 1D histogram panels, while the individual data points and 3$\sigma$ shaded contours for each sample are shown in 2D space.}
    \label{fig:stacked_corner_plots}
\end{figure*}

\begin{figure*}
    \includegraphics[width=0.49\linewidth]{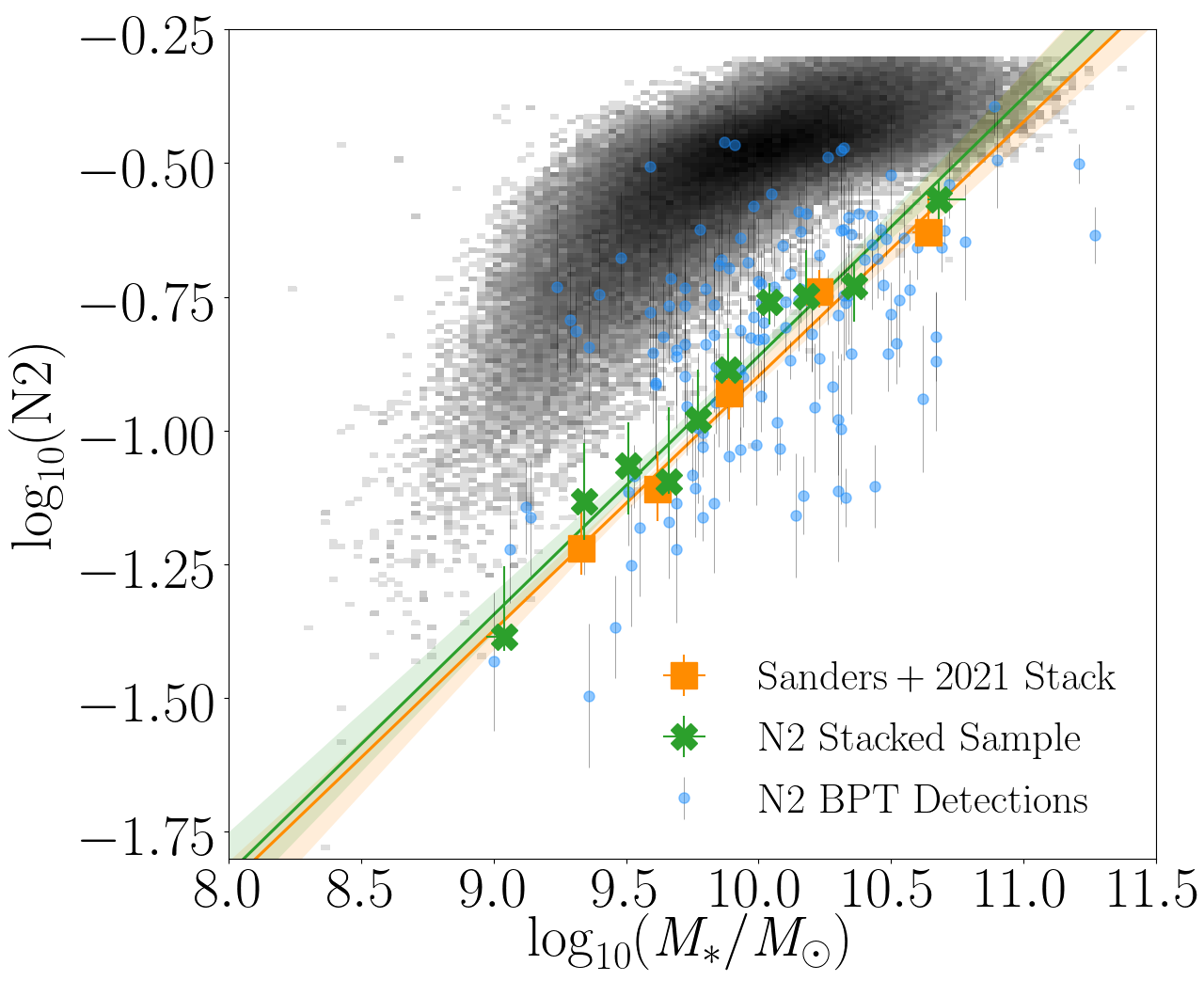}
    \includegraphics[width=0.49\linewidth]{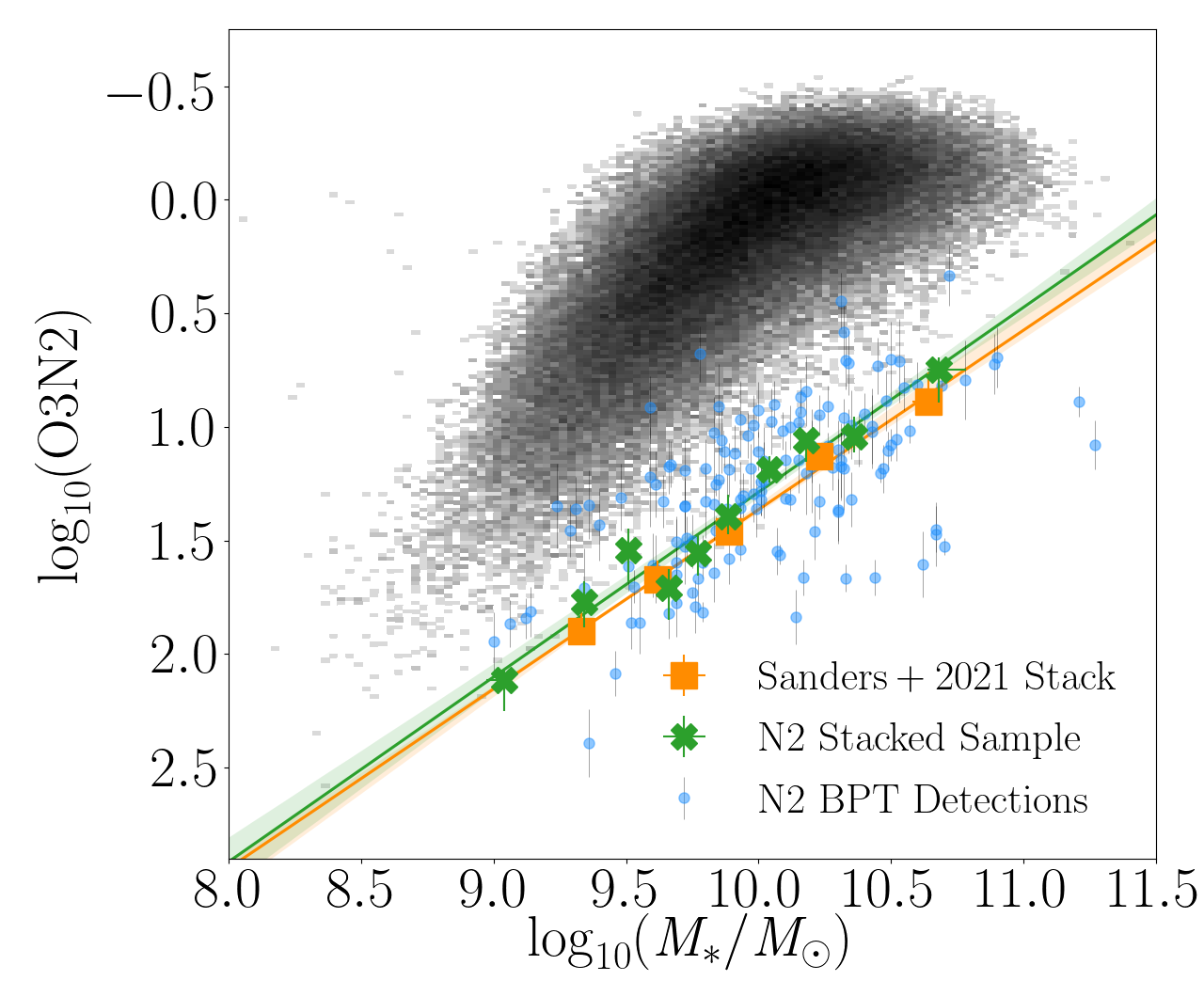}
    \includegraphics[width=0.49\linewidth]{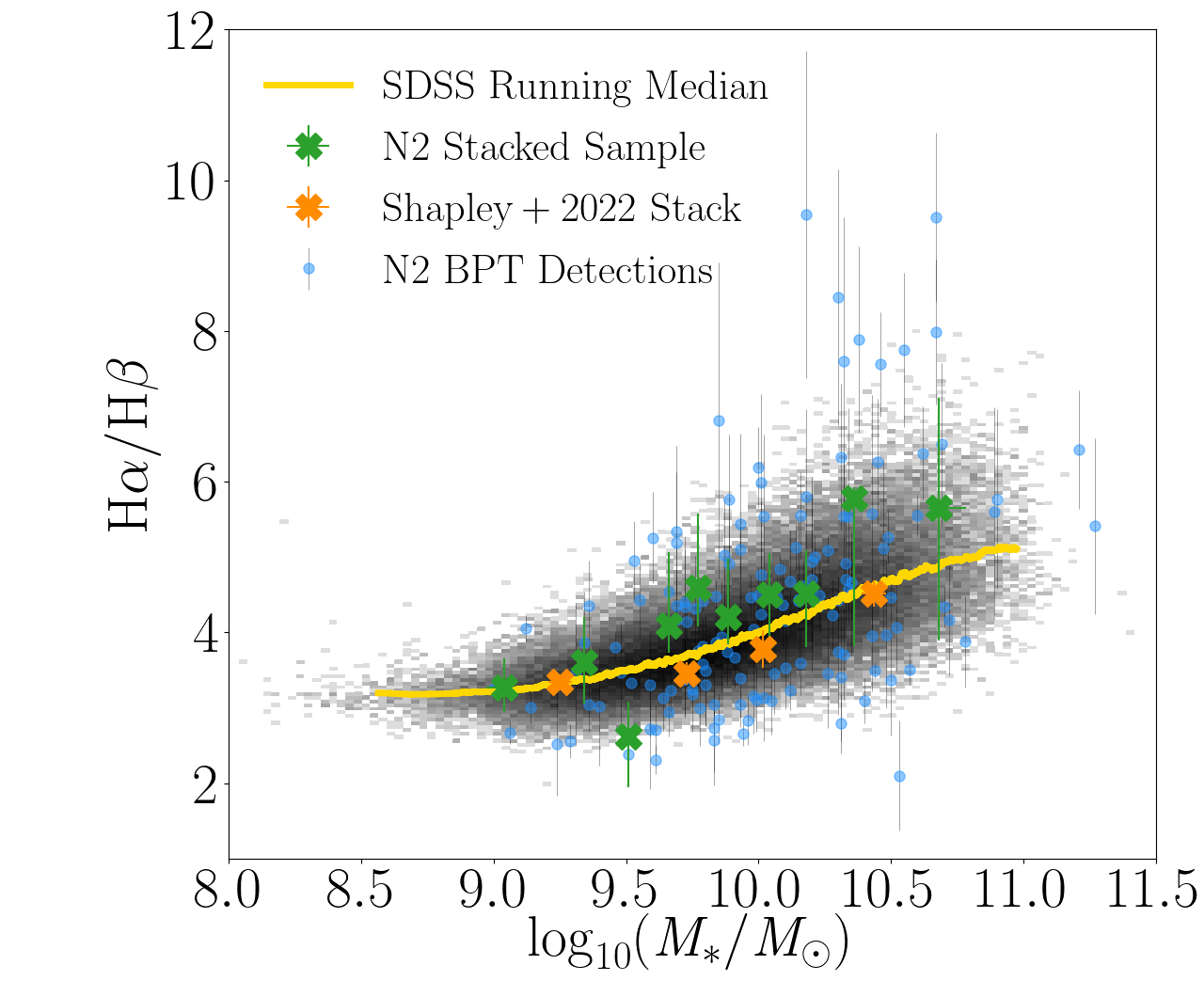}
    \caption{log$_{10}$(N2) vs. $M_{\ast}$ (upper left), log$_{10}$(O3N2) vs. $M_{\ast}$ (upper right), and H$\alpha$/H$\beta$ vs. $M_{\ast}$ (bottom). The blue points identify the 143 galaxies in Figure \ref{fig:bpt_plots} with 3$\sigma$ S/N for all relevant lines on the [N~\textsc{II}] BPT diagram. The green squares show the $z\sim2$ N2 BPT stacks. Fits to the stacks are included in the log$_{10}$(N2) vs. $M_{\ast}$ and log$_{10}$(O3N2) vs. $M_{\ast}$ panels. Also included are the MOSDEF $z\sim2$ stacks from \citet{san21} (upper left and upper right panels) and the sliding median for the MOSDEF $z\sim2$ sample from \citet{sha22} (bottom panel). The stellar mass values shown here are estimated using emission-line corrected photometry for all galaxies in the figure.}
    \label{fig:emlines_vs_mass}
\end{figure*}

\begin{figure*}
    \centering
     \subfloat[]{
       \includegraphics[width=0.32\linewidth]{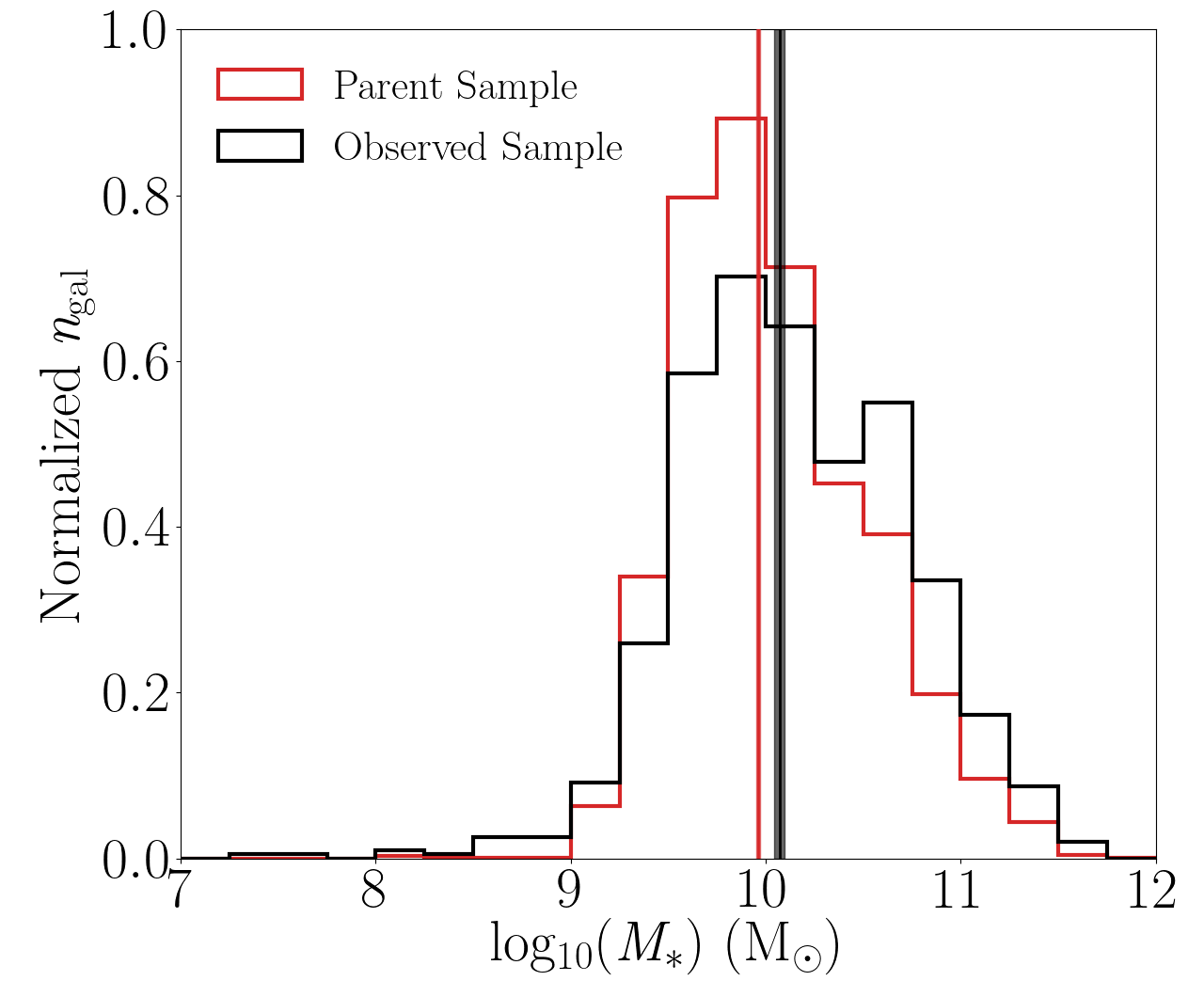}
     }
     \hfill
     \subfloat[]{
       \includegraphics[width=0.32\linewidth]{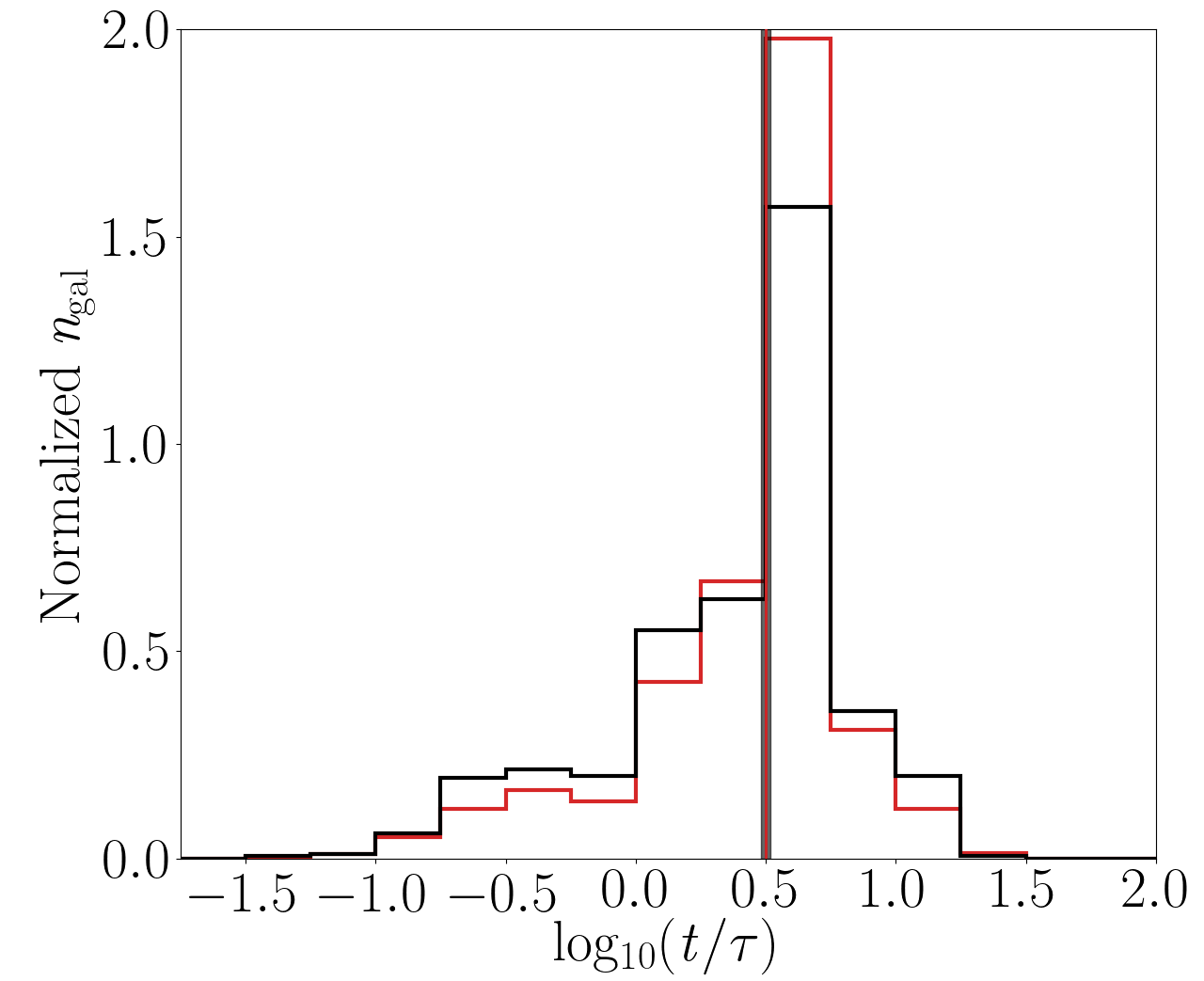}
     }
     \hfill
     \subfloat[]{
       \includegraphics[width=0.32\linewidth]{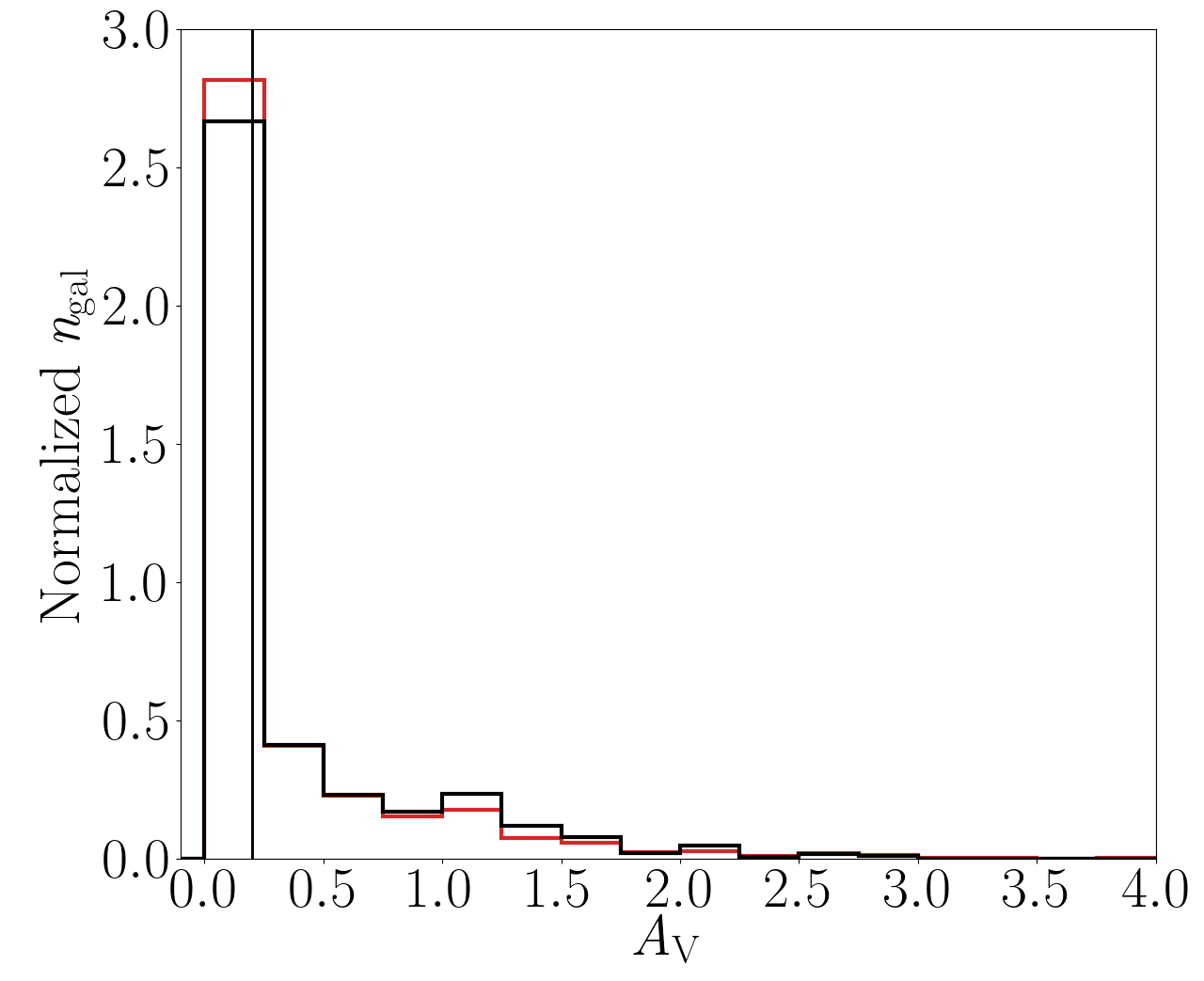}
     }
     \hfill
     \subfloat[]{
       \includegraphics[width=0.32\linewidth]{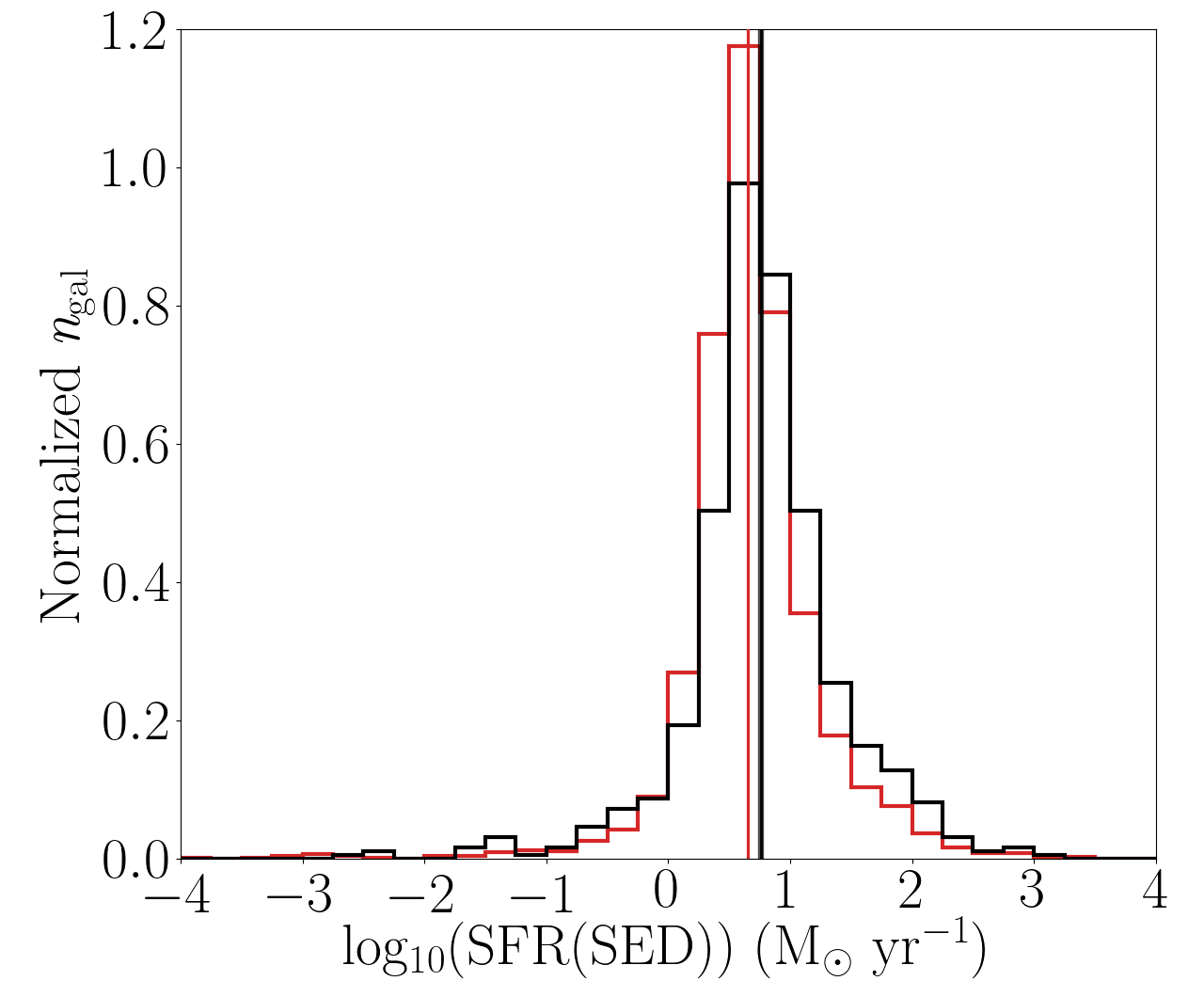}
     }
     \hfill
     \subfloat[]{
       \includegraphics[width=0.32\linewidth]{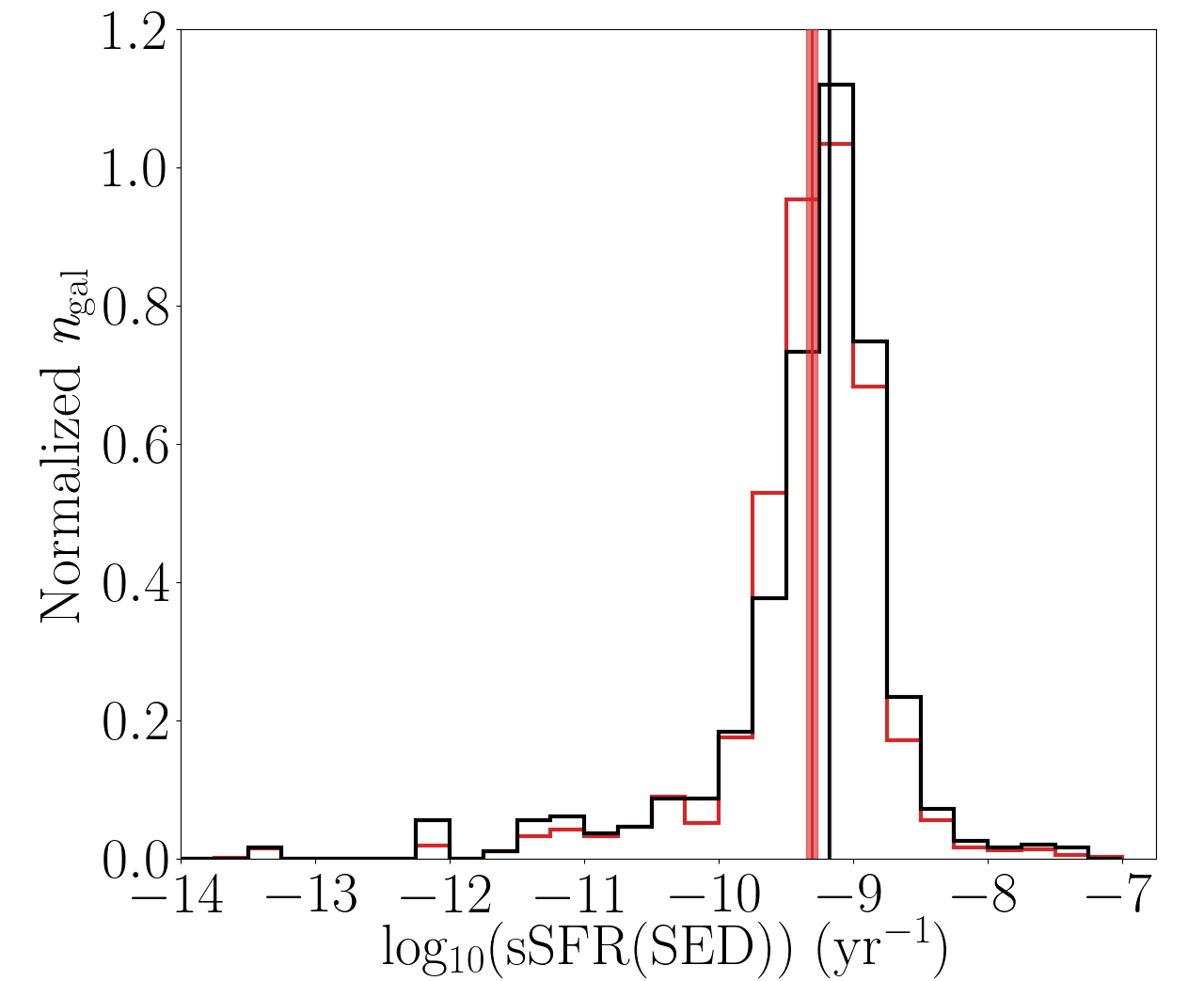}
     }
     \hfill
     \subfloat[]{
       \includegraphics[width=0.32\linewidth]{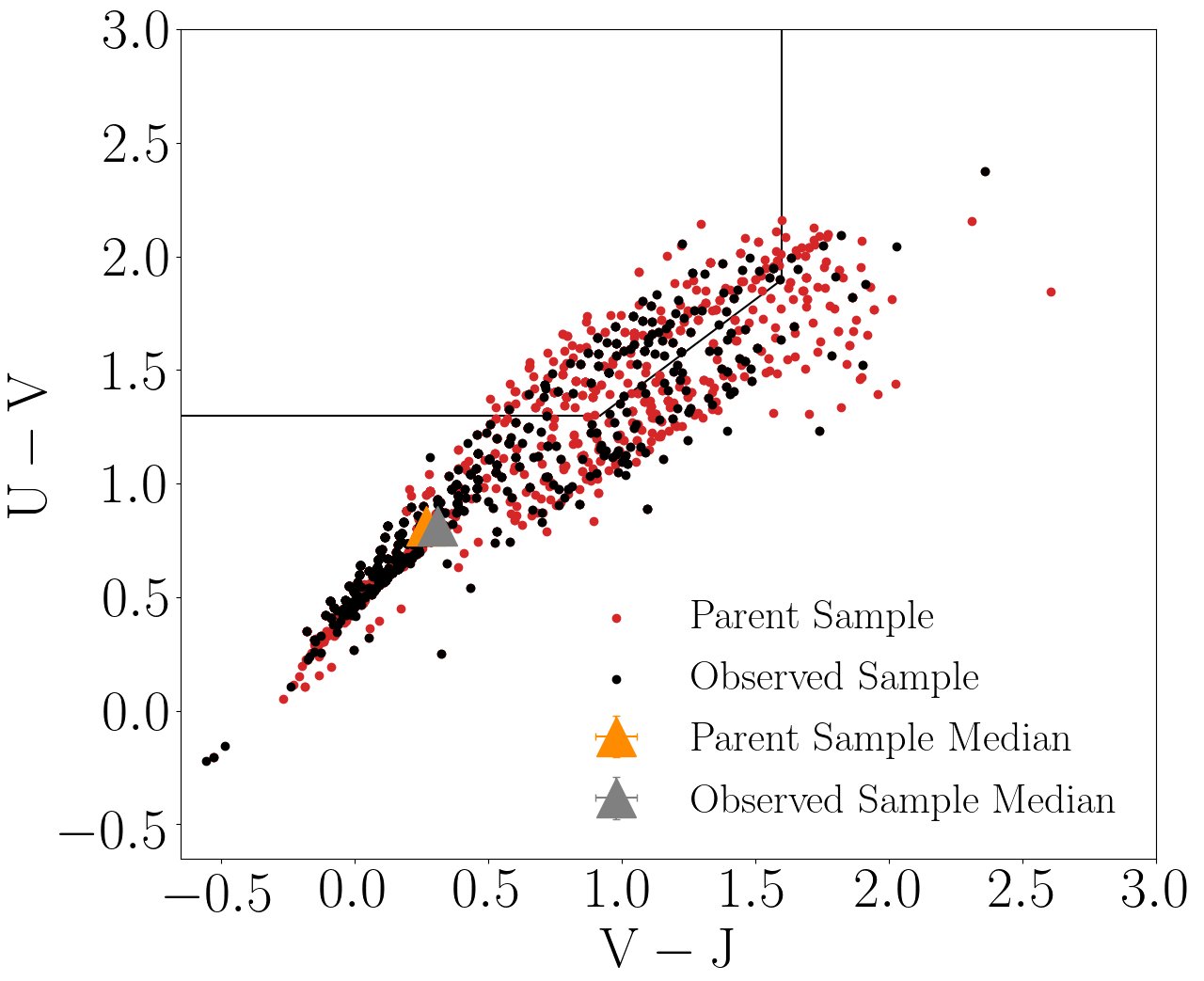}
     }
    \caption{Distribution of physical properties and sample medians with 1$\sigma$ uncertainties for the MOSDEF $z\sim2$ parent sample (red) and the $z\sim2$ observed sample (black). The values for the sample medians are given in Table \ref{tab:z2_target_parent_sample_properties}. 
    For the panels with 1D histograms, the y-axis is normalized so the area under the histogram for each sample adds up to one. 
    The distributions for the following galaxy properties are shown: (a) $M_{\ast}$, (b) log$_{10}$($t/\tau$) of the stellar population assuming a delayed-$\tau$ star formation model, (c) $A_{\rm{V}}$, (d) SFR(SED), (e) sSFR(SED), (f) the UVJ diagram. The box on the UVJ diagram separates the quiescent region (upper left) from the star-forming region (bottom half and upper right). Additionally, the sample medians on the UVJ diagram are orange ($z\sim2$ parent sample) and grey ($z\sim2$ observed sample). The galaxy properties shown here are estimated using emission-line uncorrected photometry for all galaxies in the figure. We note that the sample medians in panel (c) overlap.}
    \label{fig:z2_target_parent_sample_properties}
\end{figure*}

\begin{figure*}
    \includegraphics[width=0.49\linewidth]{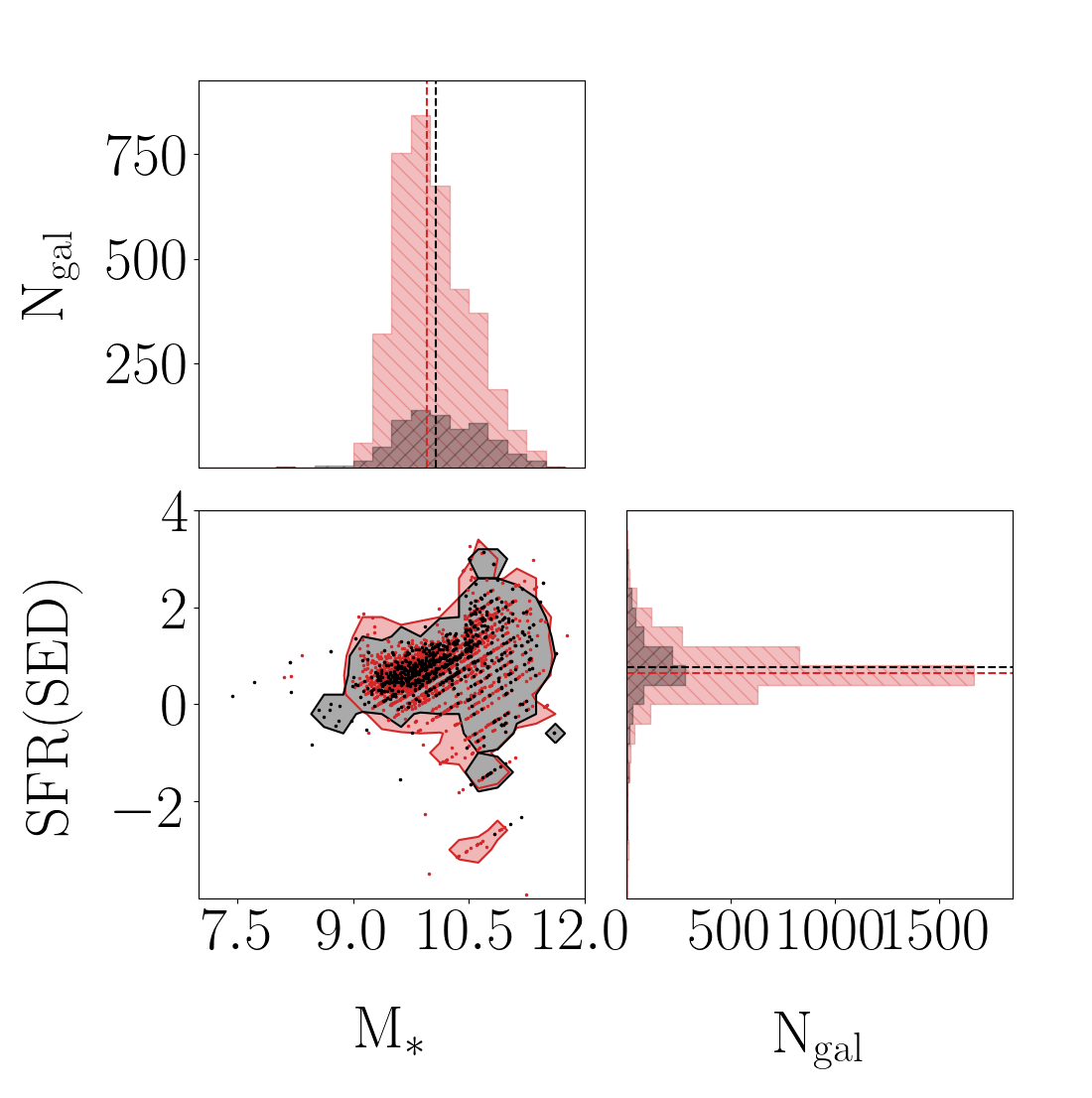}
    \includegraphics[width=0.49\linewidth]{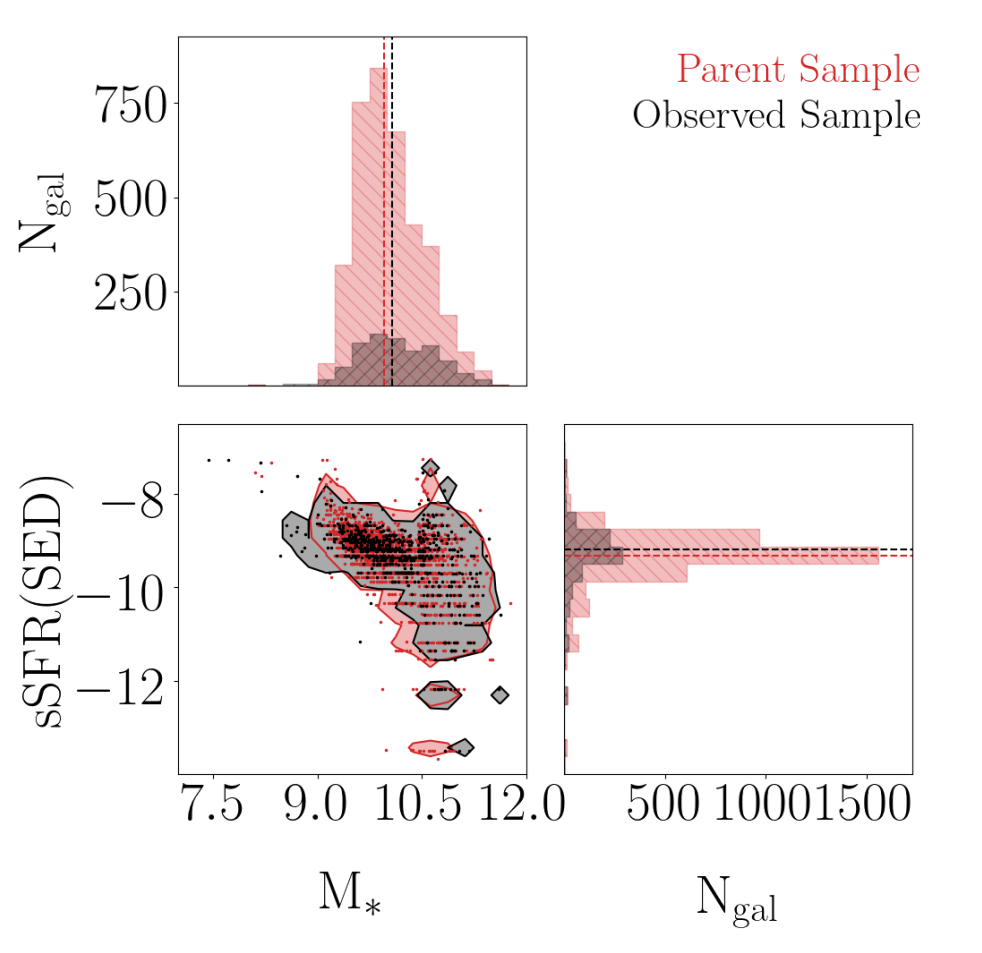}
    \includegraphics[width=0.49\linewidth]{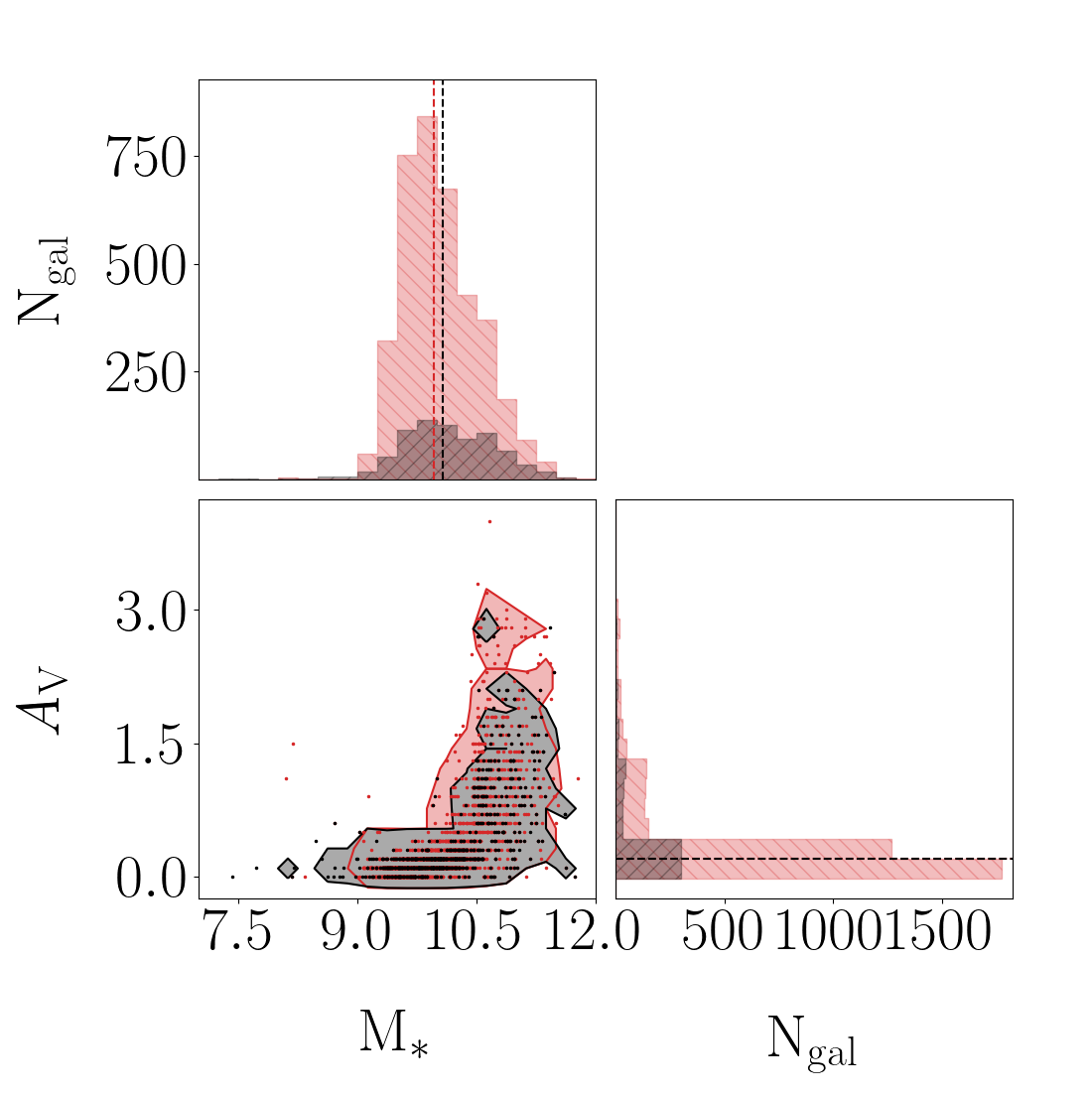}
    \includegraphics[width=0.49\linewidth]{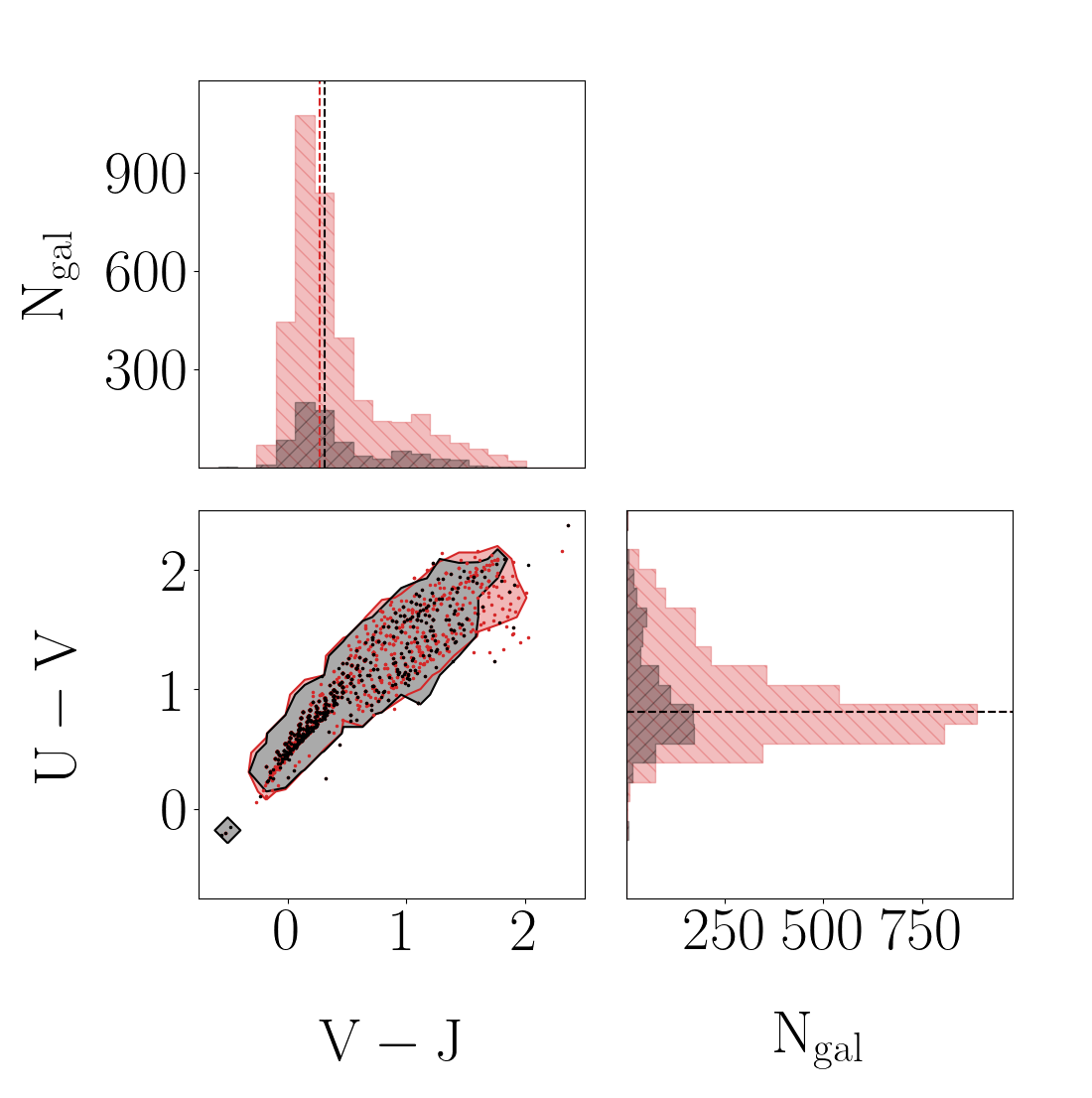}
    \caption{Corner plots for SFR(SED) . $M_{\ast}$ (upper left), sSFR(SED) vs. $M_{\ast}$ (upper right), $A_{\rm{V}}$ vs. $M_{\ast}$ (bottom left), and U$-$V vs. V$-$J (bottom right). Shown on these diagrams are the $z\sim2$ parent sample (red) and $z\sim2$ observed sample (black). The galaxy properties shown here are estimated using emission-line uncorrected photometry for all galaxies in the figure. The parameter distributions and sample medians are shown in the 1D histogram panels, while the individual data points and 3$\sigma$ shaded contours for each sample are shown in 2D space.}
    \label{fig:parent_target_corner_plots}
\end{figure*}

\begin{figure*}
    \includegraphics[width=0.49\linewidth]{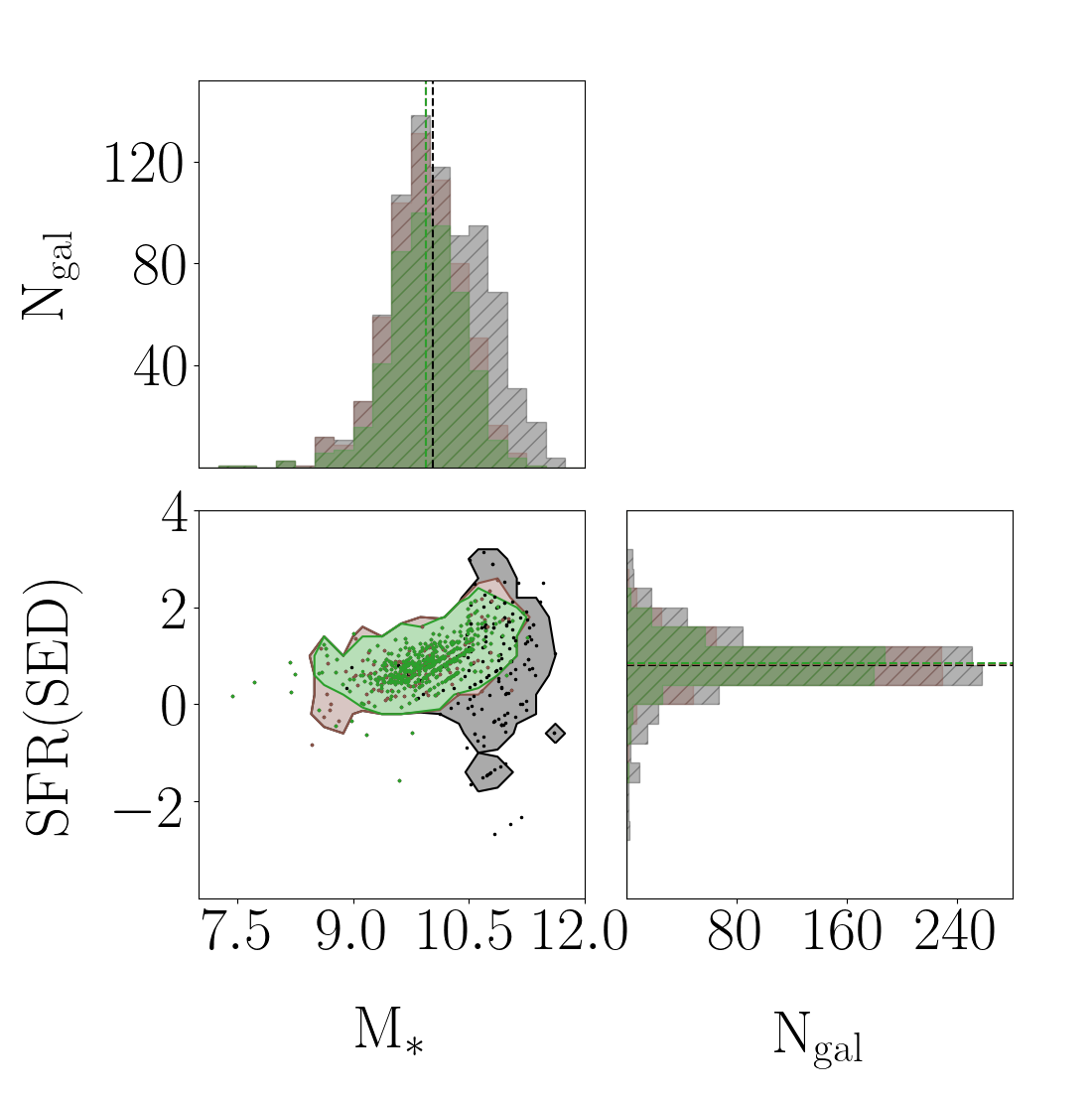}
    \includegraphics[width=0.49\linewidth]{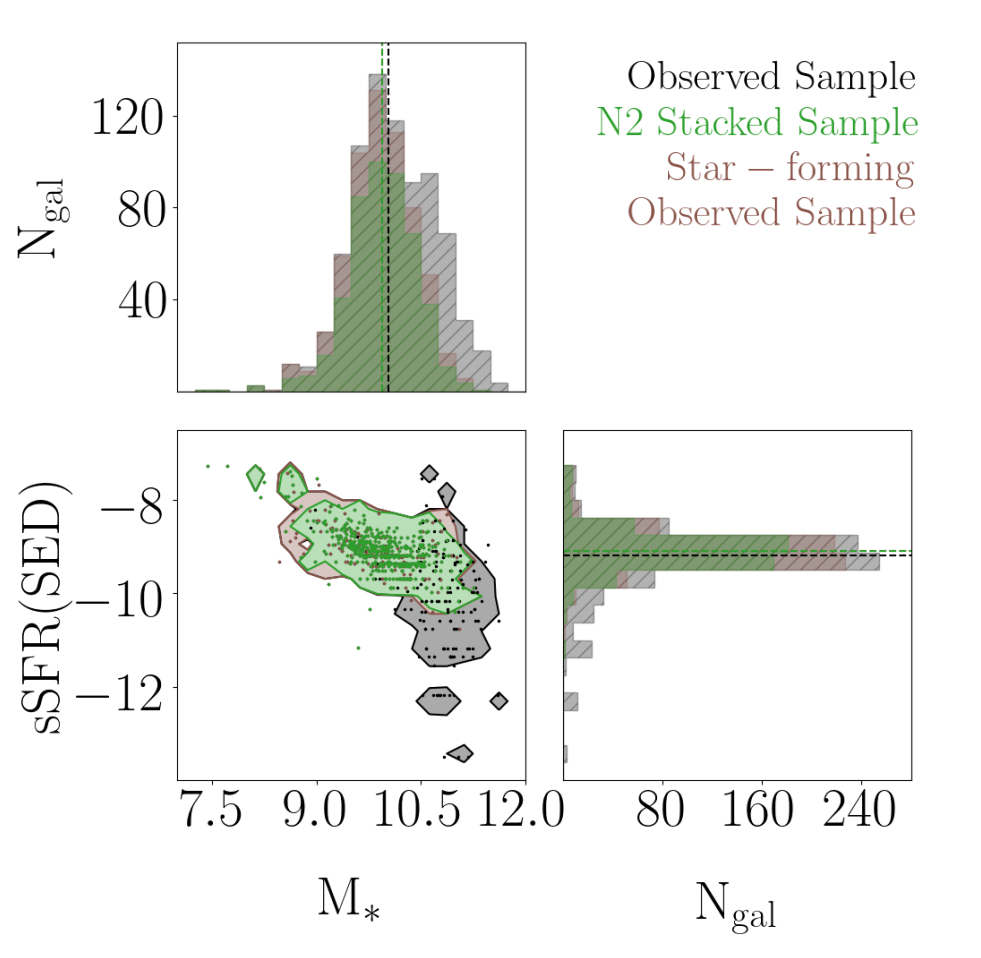}
    \includegraphics[width=0.49\linewidth]{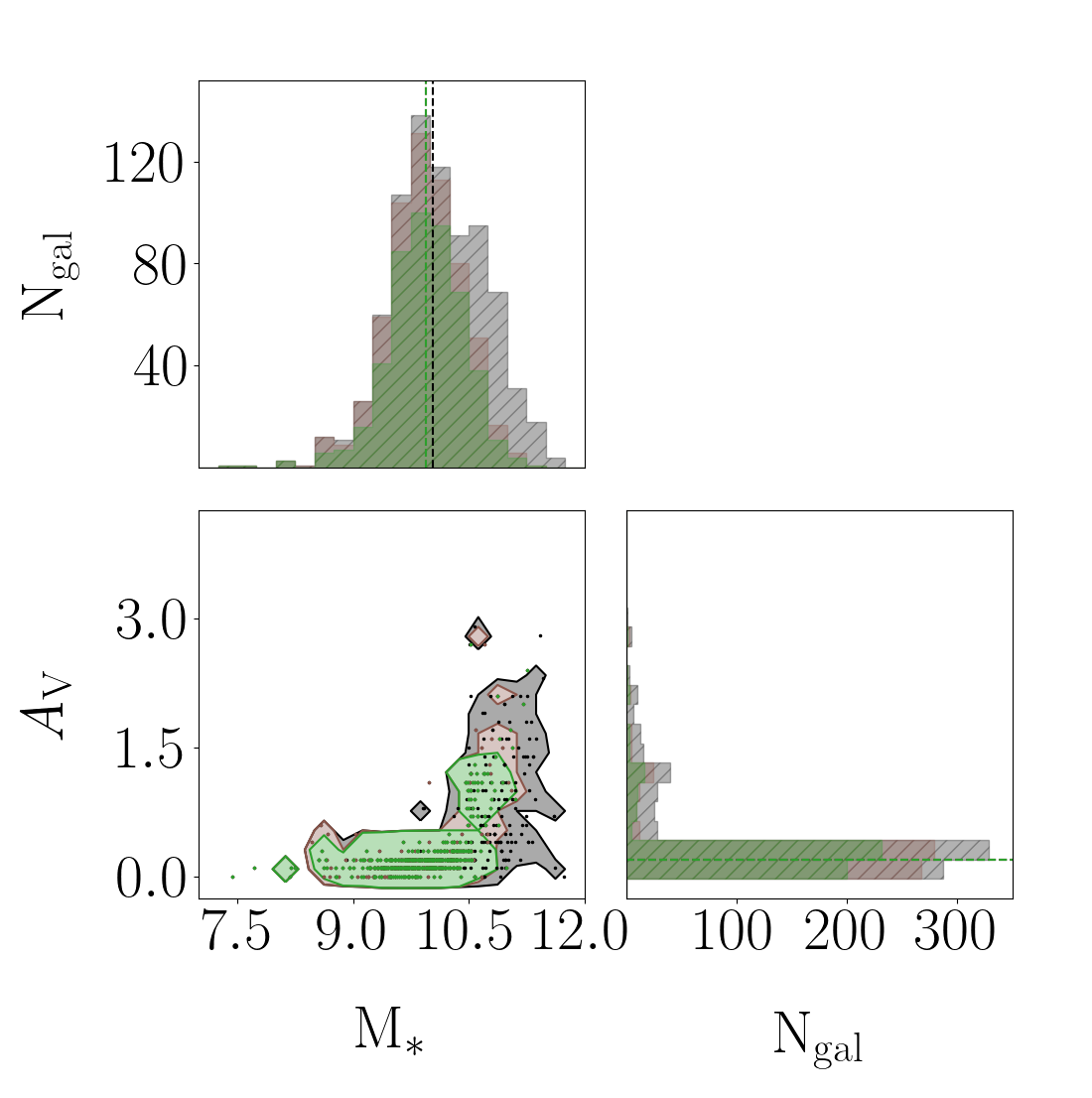}
    \includegraphics[width=0.49\linewidth]{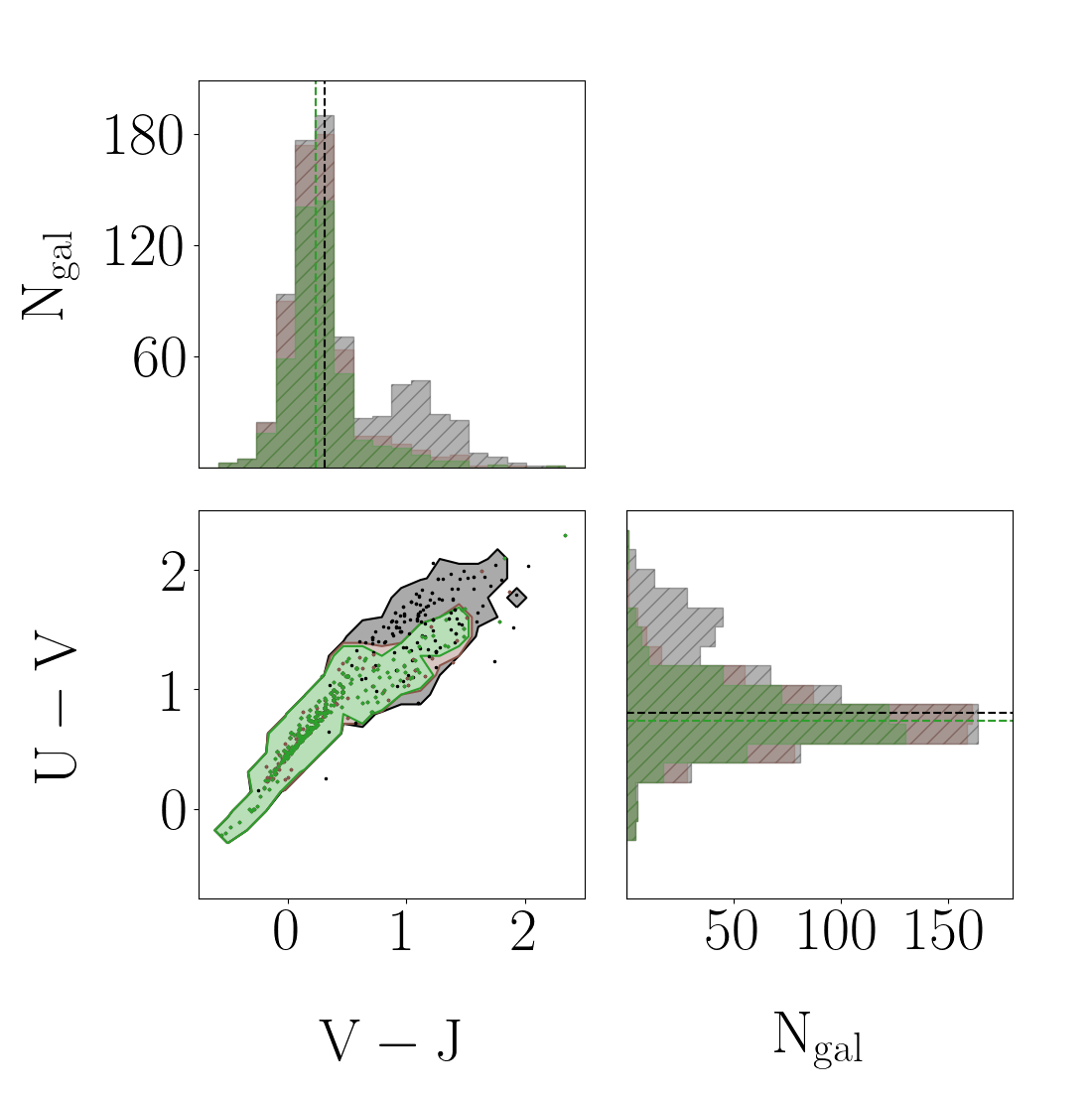}
    \caption{Corner plots for SFR(SED) vs. $M_{\ast}$ (upper left), sSFR(SED) vs. $M_{\ast}$ (upper right), $A_{\rm{V}}$ vs. $M_{\ast}$ (bottom left), and U$-$V vs. V$-$J (bottom right). Shown on these diagrams are the $z\sim2$ observed sample (black), $z\sim2$ star-forming observed sample (brown), and the $z\sim2$ N2 stacked sample (green). The galaxy properties shown here are estimated using emission-line corrected photometry for the majority of the galaxies with a MOSFIRE redshift. Otherwise, emission-line uncorrected photometry is used for the minority of galaxies without the $z_{\rm{MOSFIRE}}$. The parameter distributions and sample medians are shown in the 1D histogram panels, while the individual data points and 3$\sigma$ shaded contours for each sample are shown in 2D space.}
    \label{fig:stack_target_corner_plots}
\end{figure*}

\subsection{Physical Properties of the Stacks} \label{subsec:stack_properties}

The distributions of galaxy properties for the $z\sim2$ N2 stacked sample (478 galaxies) are shown in Figure \ref{fig:z2_stack_spec_sample_properties}. We compare this sample with the $z\sim2$ spectroscopic sample (250 galaxies) and the $z\sim2$ [N~\textsc{II}] BPT sample (143 galaxies). 
Because the total number of galaxies in each of these three samples is significantly different, we show a probability density where the area under each distribution integrates to 1 instead of raw counts to better compare the distributions of galaxies. 
Additionally, sample medians with 1$\sigma$ uncertainties are shown in Figure \ref{fig:z2_stack_spec_sample_properties} and given in Table \ref{tab:z2_stack_zspec_sample_properties}. Median uncertainties are estimated using bootstrap resampling. Specifically, we resample 10000 times and take the standard deviation of the distribution to obtain the sample median uncertainty.
We use emission-line corrected photometry to estimate these galaxy properties, since all samples in Figure \ref{fig:z2_stack_spec_sample_properties} have the MOSFIRE spectra required for estimating such corrections. 
Since the $z\sim2$ N2, S2, and O$_{32}$ stacked samples contain almost the same set of galaxies, median galaxy properties reported for the N2 stacked sample are representative for all emission-line-diagram subsets.
We also note that the median galaxy properties for the $z\sim2$ spectroscopic sample are not the identical to those reported in \citet{run22} because, as discussed in Section \ref{subsec:sed_fitting}, in this study we report galaxy properties for individual galaxies according to either the Calzetti+solar or SMC+subsolar models, based on $M_{\ast}$. In \citet{run22}, all galaxies were fit assuming the Calzetti+solar models.

\begin{table*}
    \centering
    \begin{tabular}{rrrr}
        \multicolumn{4}{c}{Median Values for Physical Properties of the MOSDEF $z\sim2$ N2 Stacked, Spectroscopic, and N2 BPT Detection Samples} \\
        \hline\hline
        Physical Property & N2 Stacked Median & Spectroscopic Median & N2 BPT Detections Median \\
        (1) & (2) & (3) & (4) \\
        \hline
 log$_{10}(M_{\ast}/M_{\odot}$) & 9.95 $\pm$ 0.03 & 10.02 $\pm$ 0.02 & 10.17 $\pm$ 0.06 \\
 log$_{10}$($t/\tau$) & 0.20 $\pm$ 0.02 & 0.15 $\pm$ 0.08 & 0.20 $\pm$ 0.04 \\
 log$_{10}$(SFR(SED)/$M_{\odot}$/yr$^{-1}$) & 0.84 $\pm$ 0.02 & 0.90 $\pm$ 0.02 & 0.95 $\pm$ 0.04 \\
 log$_{10}$(sSFR(SED)/yr$^{-1}$) & $-$9.08 $\pm$ 0.02 & $-$9.08 $\pm$ 0.03 & $-$9.17 $\pm$ 0.05 \\
 $A_{\rm{V}}$ & 0.20 $\pm$ 0.001 & 0.20 $\pm$ 0.001 & 0.20 $\pm$ 0.005 \\
 U$-$V & 0.74 $\pm$ 0.01 & 0.74 $\pm$ 0.01 & 0.80 $\pm$ 0.02 \\
 V$-$J & 0.24 $\pm$ 0.01 & 0.24 $\pm$ 0.01 & 0.31 $\pm$ 0.02 \\
        \hline
    \end{tabular}
    \caption{
  Col. (1): Physical property of the multiple MOSDEF samples shown in Figure \ref{fig:z2_stack_spec_sample_properties}. The galaxy properties shown here are estimated using emission-line corrected photometry. 
  Col. (2): Median value with uncertainty of the MOSDEF $z\sim2$ N2 stacked sample.
  Col. (3): Median value with uncertainty of the MOSDEF $z\sim2$ spectroscopic sample.
  Col. (4): Median value with uncertainty of the MOSDEF $z\sim2$ N2 BPT detection sample.}
    \label{tab:z2_stack_zspec_sample_properties}
\end{table*}

We find that the $z\sim2$ N2 stacked sample has a lower median $M_{\ast}$ and SFR(SED) compared to both the $z\sim2$ spectroscopic and N2 BPT detection samples. Additionally, the $z\sim2$ N2 stacked sample has a lower median sSFR(SED) and bluer UVJ colors compared to the $z\sim2$ N2 BPT detection sample. At the same time, the N2 stacked sample has identical median sSFR(SED) and UVJ colors compared to the $z\sim2$ spectroscopic sample. 
All other galaxy properties (i.e., $t/\tau$ and $A_{\rm{V}}$) agree within the median uncertainties for the three samples. 
The bias in $M_{\ast}$ is expected, given that galaxies with log$_{10}(M_{\ast}/M_{\odot}) < 9.0$ were removed from the $z\sim2$ spectroscopic and N2 BPT detection samples due to the documented MOSDEF incompleteness at this low mass regime. Galaxies in this mass regime were not removed from the stacked samples. 
Additionally, Figure \ref{fig:stacked_corner_plots} includes corner plots showing the 1D distributions and the 2D distributions with 3$\sigma$ contours for SFR(SED), sSFR(SED), and $A_{\rm{V}}$ vs. $M_{\ast}$ and also U$-$V vs. V$-$J colors. These diagrams show that the $z\sim2$ N2 stacked sample extends to lower $M_{\ast}$, which as previously stated is expected, but span a similar region in SFR(SED), sSFR(SED), and $A_{\rm{V}}$ to that of the $z\sim2$ spectroscopic and N2 BPT detection samples. The $z\sim2$ N2 stacked sample occupies a slightly wider range of UVJ colors (to both bluer and redder tails of 2D space) compared to the smaller high S/N samples. 
Combining these results with the emission-line analysis in Section \ref{subsec:stack_emlines} reveals that these differences in host galaxy properties between the $z\sim2$ N2 stacked sample and $z\sim2$ N2 detection sample do not strongly affect the emission-line ratios. 

Figure \ref{fig:emlines_vs_mass} compares the relationships between N2, O3N2, and Balmer decrement (i.e., H$\alpha$/H$\beta$) vs. $M_{\ast}$ for different MOSDEF subsamples. These plots include the detections and the $z\sim2$ N2 stacked sample from the [N~\textsc{II}] BPT diagram in the left panel of Figure \ref{fig:bpt_plots}. 
Also included on all three diagrams is the SDSS sample. 
For the N2 and O3N2 vs. $M_{\ast}$ panels (top right and top left, respectively), we also include a fit to the $z\sim2$ N2 stacked sample and the $z\sim2$ MOSDEF stacks (with fit) from \citet{san21}. 
The stacked sample from \citet{san21} contains 280 star-forming galaxies at $z\sim2$ with S/N$_{[\rm{O~\textsc{III}}]\lambda5008} \geq 3$ and coverage of [O~\textsc{II}]$\lambda\lambda$3727,3730, [Ne~\textsc{II}]$\lambda$3870, and H$\beta$. There are four bins within the range $9.0 \leq \rm{log}_{10}(M_{\ast}/M_{\odot}) \leq 10.5$ and one additional bin incorporating galaxies with log$_{10}(M_{\ast}/M_{\odot}) > 10.5$. Galaxies with log$_{10}(M_{\ast}/M_{\odot}) < 9.0$ are omitted from the sample. These mass cuts are made because the range of $M_{\ast}$ = 10$^{9.0-10.5}$ is where the MOSDEF survey has the highest spectroscopic success rate \citep{kri15}. 

In the N2 vs. $M_{\ast}$ and O3N2 vs. $M_{\ast}$ panels, we find that the linear regressions fit to the $z\sim2$ N2 stacked sample and \citet{san21} stacks overlap within the uncertainties, which suggests that the smaller MOSDEF subsample with high S/N is not significantly biased with respect to the full MOSDEF star-forming sample. 
Line-ratio measurements from both the larger $z\sim2$ stacked samples constructed here and those presented in \citet{san21} are offset from the local SDSS star-forming sample, indicating evolution between $z\sim2$ and $z\sim0$ in these metallicity-sensitive rest-optical emission-line ratios at fixed $M_{\ast}$. 

For the H$\alpha$/H$\beta$ vs. $M_{\ast}$ diagram, we also include the sliding medians from the $z\sim2$ MOSDEF sample from \citet{sha22}. This sample is very similar to the spectroscopic sample from \citet{san21}, which required $\geq3\sigma$ detections for [O~\textsc{II}]$\lambda\lambda$3727,3730, [Ne~\textsc{II}]$\lambda$3870, and H$\beta$, and [O~\textsc{II}]$\lambda$5008 (i.e., more restrictive than the stacked sample described above).  \citet{sha22} adopted this sample from \citet{san21}, with the added requirement of S/N$_{\rm{H}\alpha} \geq 3$, in order to probe dust attenuation. 
The $z\sim2$ N2 stacked sample shows a slightly higher H$\alpha$/H$\beta$ ratio on average compared to the \citet{sha22} sliding median bins. This offset is not significant, as the $z\sim2$ N2 stacked sample bins overlap with the \citet{sha22} sliding median bins within the uncertainties at fixed $M_{\ast}$. 
Both $z\sim2$ MOSDEF samples intersect with the running median for the SDSS $z\sim0$ star-forming sample, indicating that there is no significant evolution in the Balmer decrement at fixed $M_{\ast}$. 
Many studies have found no evolution in the dust attenuation at fixed $M_{\ast}$ between $z\sim2$ and $z\sim0$, using a wide variety of methods to trace dust attenuation: the Balmer decrement (e.g., \citealt{dom13, kas13, pri14, sha22}), the magnitude of far-UV 1600 \AA \ attenuation (e.g., \citealt{pan15, mcl18}), the fraction of obscured star formation (e.g., \citealt{whi17}), and the ratio of far-IR to UV SFRs (e.g., \citealt{meu99, bou16}).

\begin{table*}
    \centering
    \begin{tabular}{rrrrr}
        \multicolumn{5}{c}{Median Values for Physical Properties of the MOSDEF $z\sim2$ Parent and Observed Samples} \\
        \hline\hline
        Physical Property & Parent Median & Parent 68\% Confidence Interval & Observed Median & Observed 68\% Confidence Interval \\
        (1) & (2) & (3) & (4) & (5) \\
        \hline
 log$_{10}(M_{\ast}/M_{\odot}$) & 9.96 $\pm$ 0.01 & [9.77, 10.22] & 10.07 $\pm$ 0.03 & [9.85, 10.43] \\
 log$_{10}$($t/\tau$) & 0.50 $\pm$ 0.00 & [0.40, 0.60] & 0.50 $\pm$ 0.02 & [0.30, 0.60] \\
 log$_{10}$(SFR(SED)/$M_{\odot}$/yr$^{-1}$) & 0.65 $\pm$ 0.01 & [0.50, 0.83] & 0.76 $\pm$ 0.02 & [0.57, 0.96] \\
 log$_{10}$(sSFR(SED)/yr$^{-1}$) & $-$9.31 $\pm$ 0.04 & [$-$9.41, $-$9.08] & $-$9.18 $\pm$ 0.01 & [$-$9.36, $-$9.04] \\
 $A_{\rm{V}}$ & 0.20 $\pm$ 0.001 & [0.10, 0.20] & 0.20 $\pm$ 0.000 & [0.10, 0.30] \\
 U$-$V & 0.81 $\pm$ 0.01 & [0.71, 1.00] & 0.81 $\pm$ 0.01 & [0.71, 1.03] \\
 V$-$J & 0.27 $\pm$ 0.01 & [0.17, 0.43] & 0.31 $\pm$ 0.01 & [0.18, 0.51] \\
        \hline
    \end{tabular}
    \caption{
  Col. (1): Physical property of the multiple MOSDEF samples shown in Figure \ref{fig:z2_target_parent_sample_properties}. The galaxy properties shown here are estimated using emission-line uncorrected photometry. 
  Col. (2): Median value with uncertainty of the MOSDEF $z\sim2$ parent sample.
  Col. (3): The lower and upper values of the 68\% confidence interval for the MOSDEF $z\sim2$ parent sample.
  Col. (4): Median value with uncertainty of the MOSDEF $z\sim2$ observed sample.
  Col. (5): The lower and upper values of the 68\% confidence interval for the MOSDEF $z\sim2$ observed sample.}
    \label{tab:z2_target_parent_sample_properties}
\end{table*}

\subsection{Comparison of the $z\sim2$ Parent and Observed Samples} \label{subsec:parent_vs_targeted_samples}

We have shown that the subset of MOSDEF $z\sim2$ star-forming galaxies with high S/N spectra is representative of the stacked sample of MOSDEF $z\sim 2$ star-forming galaxies with coverage of the [N~\textsc{II}] BPT, [S~\textsc{II}] BPT, and O$_{32}$ vs. R$_{23}$ diagrams. 
Now we turn our attention to whether the full sample of $z\sim2$ galaxies observed in MOSDEF (including both star-forming and non-star-forming targets) is reflective of the complete catalog of galaxies that could have been targeted by MOSDEF. 
Figure \ref{fig:z2_target_parent_sample_properties} shows the 1D distributions of the galaxy properties, with sample medians overlaid, for the $z\sim2$ parent and observed samples. 
As in Figure~\ref{fig:z2_stack_spec_sample_properties}, we plot the histograms as probability densities instead of raw galaxy counts to best compare the distribution of galaxies. 
Table \ref{tab:z2_target_parent_sample_properties} lists the median galaxy properties of the two samples. 
Again, the 1$\sigma$ uncertainties on the medians are estimated using bootstrap resampling. 
In this comparison of the properties of the $z\sim 2$ parent and observed samples, we use photometry that has not been corrected by emission-lines (unlike what is done in Section \ref{subsec:stack_properties}) because we do not have a MOSFIRE spectrum for the majority of the $z\sim2$ parent sample. As we state in Section \ref{subsec:sed_fitting}, emission-lines can bias the modeling to favor older stellar population ages. Therefore, for a uniform analysis, we use uncorrected photometry for all galaxies in the $z\sim2$ parent and observed samples. 

We find that the $z\sim2$ parent sample has a lower median $M_{\ast}$, SFR(SED), sSFR(SED) than the $z\sim2$ observed sample. However, the difference in the sample medians is only $\sim$0.1 dex. Table \ref{tab:z2_target_parent_sample_properties} also includes the lower and upper values of the 68\% confidence interval for the distributions, with only small differences between the two samples. Given that that the majority of galaxies in both samples span $\sim$4 orders of magnitude in each of these parameter spaces (with outlier galaxies extending even farther), the observed median differences are minor. Additionally, the $z\sim2$ parent sample has a slightly bluer median V$-$J color compared to the $z\sim2$ observed sample. All other properties (i.e., $t/\tau$, $A_{\rm{V}}$, and U$-$V color) agree within the median uncertainties. 

Figure \ref{fig:parent_target_corner_plots} includes corner plots of SFR(SED), sSFR(SED), and $A_{\rm{V}}$ vs. $M_{\ast}$, which show 1D histograms with sample median of each parameter overlaid, and the 2D distribution of data points outlined by a 3$\sigma$ contour. The $z\sim2$ parent and observed samples mostly occupy similar regions in 2D space in all three of these diagrams, with only minor differences at the tails of the distributions (e.g., the $z\sim2$ parent sample extends to higher $A_{\rm{V}}$ and lower SFR(SED) compared to the $z\sim2$ observed sample). 

In summary, Figures \ref{fig:z2_target_parent_sample_properties} and \ref{fig:parent_target_corner_plots} show that the sample of $z\sim2$ galaxies actually observed with MOSFIRE as a part of the MOSDEF survey is representative of the $z\sim2$ parent sample from which it was drawn. We discuss the significance of this result in Section \ref{sec:discussion}.

\section{Discussion} \label{sec:discussion}

In this section, we discuss the completeness of the MOSDEF $z\sim2$ survey from multiple angles. Section \ref{subsec:mosdef_star_forming_galaxy_completeness} considers the implications of the completeness of the $z\sim2$ spectroscopic sample with respect to the larger group of star-forming galaxies in the $z\sim2$ stacked samples.
Section \ref{subsec:mosdef_targeted_completeness} expands the discussion from Section \ref{subsec:mosdef_star_forming_galaxy_completeness}.

\subsection{MOSDEF $z\sim2$ Star-forming Galaxy Completeness} \label{subsec:mosdef_star_forming_galaxy_completeness}

In Sections \ref{subsec:stack_emlines} and \ref{subsec:stack_properties}, we show that the emission-line and physical properties of the $z\sim2$ stacked samples and the subsets of the stacked samples with high S/N spectra are very similar. 
The small differences in median properties are minor given the large range of values seen in the distributions. 
The fact that the smaller spectroscopic samples are representative of the full star-forming catalog in the MOSDEF survey is significant, as we can create composite spectra using stacking to increase the available sample size in future studies. 

Additionally, the agreement with the \citet{san21} sample on the N2 and O3N2 vs. $M_{\ast}$ diagrams and with the \citet{sha22} sample on the H$\alpha$/H$\beta$ vs. $M_{\ast}$ diagram shows that the results and conclusions based on those smaller samples are representative of the larger $z\sim2$ star-forming population. 
The offset between local and $z\sim2$ samples on the N2 and O3N2 vs. $M_{\ast}$ diagrams reflects the well-known evolution of the mass-metallicity relation, MZR, (e.g., \citealt{and13, ste14, san15}). 

\citet{sha22} discuss that the lack of evolution in dust attenuation at fixed $M_{\ast}$ points to an evolution towards increasing patchiness in dust attenuation at higher redshift, given the elevated $M_{\rm{dust}}/M_{\ast}$ ratios inferred at earlier times \citep[e.g.,][]{don20,mag20,shi22}. Alternatively, this lack of evolution in attenuation could be explained by evolution of the wavelength-dependent dust mass absorption coefficient ($\kappa_{\lambda}$). 
This dust cross-section per unit dust mass ($\kappa_{\lambda}$) has been shown to be inversely correlated with gas surface density, $\Sigma_{\rm{gas}}$ \citep{cla19}, in the local universe. At $z\sim2$, $\Sigma_{\rm{gas}}$ is shown to be significantly higher than at $z\sim 0$. If the anticorrelation between  $\kappa_{\lambda}$ and gas surface density holds at $z\sim 2$, typical $\kappa_{\lambda}$ values may be lower at high redshift, compared to what is observed in the local universe. 
Another possibility could be a steeper slope between $M_{\rm{dust}}/M_{\ast}$ and gas-phase oxygen abundance; however, \citet{sha20,pop22} find no evolution in this relationship out to $z\sim 2$ at near-solar metallicities. One caveat is that these studies investigated a narrow mass range (log$_{10}(M_{\ast}/M_{\odot}) \sim 10.5-11$). 
In summary, we confirm the conundrum from \citet{sha22} with a larger and more complete data set. 
Additional ALMA observations of dust continuum and CO emission from $z\sim 2$ star-forming galaxies spanning a wider range of $M_{\ast}$ and metallicity is required for understanding this intriguing lack of evolution in dust attenuation at fixed stellar mass over 10 billion years.

\subsection{MOSDEF $z\sim2$ Observed Sample Completeness} \label{subsec:mosdef_targeted_completeness}

In this study we show that the high S/N $z\sim2$ detection samples analyzed in previous MOSDEF works are representative of the larger $z\sim2$ stacked samples analyzed for the first time here. Furthermore, the MOSDEF $z\sim2$ observed sample is representative of the full $z\sim2$ parent sample. However, in \citet{run22}, we show that the $z\sim2$ spectroscopic sample containing only star-forming galaxies suffers from incompleteness in the high-mass, red-UVJ color regime compared to the $z\sim2$ observed sample. 
It is not clear if this incompleteness is relevant for our study of $z\sim2$ \textit{star-forming} galaxies, given that it is dominated by the inclusion of AGN and quiescent galaxies. Therefore, in this section, we investigate if incompleteness also exists between the $z\sim2$ stacked sample and $z\sim2$ observed sample, but for star-forming galaxies alone. 
To do so we look at a new subset of the $z\sim2$ observed sample, from which all AGN and quiescent galaxies have been removed. Of the 786 galaxies in the MOSDEF $z\sim2$ observed sample, 171 galaxies are removed for either containing an AGN and/or fall into the quiescent box as defined in \citet{wil09}. This 615-galaxy sample is less restrictive than the $z\sim2$ stacked samples in that we do not require coverage of any specific set of emission-lines and also do not require a robust $z_{\rm{MOSFIRE}}$. We will hereafter refer to this 615 galaxy sample as the ``$z\sim2$ star-forming observed sample.''

Figure \ref{fig:stack_target_corner_plots} includes corner plots of SFR(SED), sSFR(SED), and $A_{\rm{V}}$ vs. $M_{\ast}$ and also U$-$V vs. V$-$J colors. The diagrams feature 1D histograms with sample medians overlaid for each parameter as well as the 2D distribution of data points traced by a 3$\sigma$ contour. 
We find that the $z\sim2$ N2 stacked sample is incomplete with respect to the full $z\sim2$ observed sample, as the latter extends to higher $M_{\ast}$, lower SFR(SED) and sSFR(SED), and redder UVJ colors. These results reflect what was found in \citet{run22}.
On the other hand, the $z\sim2$ N2 stacked sample is very similar to the $z\sim2$ star-forming observed sample, with the 3$\sigma$ contours for both samples covering almost identical ranges in all four diagrams. 

This analysis shows that the sample of AGN and quiescent galaxies observed by MOSDEF caused the incompleteness between the $z\sim2$ spectroscopic and observed samples reported in \citet{run22}. These galaxies make up the population of very red, high mass galaxies with little star formation in the $z\sim2$ observed sample. This combination of properties is not common for galaxies actively forming stars. In conclusion, the high S/N $z\sim2$ detection samples of star-forming galaxies used in MOSDEF emission-line studies to date are representative of the full set of $z\sim2$ star-forming galaxies in the MOSDEF parent catalog, and, therefore, the rest-optical-magnitude-limited star-forming galaxy population at $z\sim 2$.

\section{Summary} \label{sec:summary}

In this study we investigate the completeness of the $z\sim2$ MOSDEF survey. We compare the different subsets of star-forming galaxies with high S/N rest-optical emission-line detections (as described in Table \ref{tab:mosdef_samples}) against composite spectra for the full star-forming sample created using spectral stacking. We use a combination of SED and emission-line fitting to test if the high S/N subset typically used in previous MOSDEF studies is representative of the broader population of star-forming galaxies. 
Additionally, we compare the full $z\sim2$ sample observed by MOSDEF (including AGN and quiescents) with all galaxies in the five 3D-HST fields targeted by MOSDEF (AEGIS, COSMOS, GOODS-N, GOODS-S, and UDS) at $1.9 \leq z \leq 2.7$ that meet the MOSDEF $H$-band magnitude selection criteria ($H_{\rm{AB}} =$ 24.5). In this comparison, we utilize SED fitting to investigate if the properties of the galaxies observed by MOSDEF are representative of the $z\sim2$ universe. 

The main results are as follows:
\begin{enumerate}
    \item The sequence of bins in the $z\sim2$ N2, S2, and O$_{32}$ stacked sample occupy similar regions of the [N~\textsc{II}] BPT, [S~\textsc{II}] BPT, and O$_{32}$ vs. R$_{23}$ diagrams to those of the smaller $z\sim2$ detection samples with high S/N spectra. Additionally, the bins in the $z\sim2$ N2 and S2 stacked samples overlap the binned medians for the $z\sim2$ N2 BPT and S2 BPT detection samples from \citet{run22}. On the [N~\textsc{II}] BPT diagram, the lowest-$M_{\ast}$ bin in the $z\sim 2$ N2 stacked sample extends the sequence to higher N2 and/or O3. In addition, the $z\sim2$ N2 stacked sample has the same median offset (within the uncertainties) from the SDSS local sequence as the binned medians from \citet{run22} (0.10 $\pm$ 0.04 dex and 0.12 $\pm$ 0.02 dex, respectively). The $z\sim2$ N2 stacked sample overlaps with the fit to early MOSDEF data from \citet{sha15}. 
    This agreement between the high S/N $z\sim2$ detection samples and the larger stacked samples utilizing all star-forming galaxies at $z\sim2$ suggests that the emission-lines from smaller high S/N subsets typically used in MOSDEF studies are not biased with respect to the sample from which they are drawn.
    \item We find an offset between the $z\sim2$ N2 stacked sample and the local star-forming sequence on diagrams using emission-line ratios plotted vs. $M_{\ast}$ as proxies for the mass-metallicity relationship. These emission-line ratios include N2 and O3N2. The magnitude of the offset between $z\sim 2$ and $z\sim 0$ line ratio at fixed $M_{\ast}$ is similar to that derived from previous studies using high S/N MOSDEF samples. 
    We do not find evolution in dust attenuation (traced by H$\alpha$/H$\beta$) between $z\sim2$ and local $z\sim0$ galaxies. 
    \item The MOSDEF $\sim2$ observed sample has a higher median $M_{\ast}$, SFR(SED), and sSFR(SED) and redder median V$-$J color compared to the $z\sim2$ parent sample. However, the offset in median $M_{\ast}$, SFR(SED), and sSFR(SED) between the two samples is relatively small ($\sim$0.1 dex) considering that the distributions of these parameters span more than four orders of magnitude. This indicates that the galaxies at $z\sim2$ observed by the MOSDEF team are not biased from the full galaxy population at this epoch.
    \item Removing AGN and quiescent galaxies from the $z\sim2$ observed sample resolves the issue of incompleteness between the  $z\sim2$ observed sample and the $z\sim2$ spectroscopic sample reported in \citet{run22}. The $z\sim2$ spectroscopic sample has very similar galaxy properties to the full catalog of $z\sim2$ star-forming galaxies in the MOSDEF survey. 
\end{enumerate}

The lack of bias between the $z\sim2$ stacked sample and the smaller high S/N subsets indicate that, moving forward, we can use the latter in future studies as a robust proxy for the $z\sim 2$ star-forming galaxy population. 
This lack of bias also implies that past results on excitation and metallicity for emission-line selected MOSDEF samples are representative of the $z\sim2$ star-forming galaxies.
Large statistical samples are needed to better understand the proper utilization of emission-line ratios as calibrations for gas-phase oxygen abundance patters. Such studies are essential and will push us closer to create relationships (e.g., MZR, FMR) in the $z\sim2$ universe approaching what already exists locally. 
The construction of composite spectra using spectral stacking has shown to be a valuable tool in assembling these larger samples. Faint and dusty galaxies require longer integration times to capture the full set of rest-optical emission-lines. For these extreme galaxies it can be difficult to obtain 3$\sigma$ detections for bright lines (e.g., H$\alpha$) on an individual basis. In these situations, measuring weaker lines (e.g., [N~\textsc{II}]$\lambda$6585) is significantly more challenging. Incorporating these galaxies into composite spectra is essential for creating samples that are more representative of the $z\sim2$ universe.

\section*{Acknowledgements}
We acknowledge support from NSF AAG grants AST1312780, 1312547, 1312764, 1313171, 2009313, 2009085, 2009278, grant AR-13907 from the Space Telescope Science Institute, and grant NNX16AF54G from the NASA ADAP program. We also acknowledge a NASA contract supporting the ``WFIRST Extragalactic Potential Observations (EXPO) Science Investigation Team'' (15-WFIRST15-0004), administered by GSFC. Support for this work was also provided through the NASA Hubble Fellowship grant \#HST-HF2-51469.001-A awarded by the Space Telescope Science Institute, which is operated by the Association of Universities for Research in Astronomy, Incorporated, under NASA contract NAS5-26555.
We thank the 3D-\textit{HST} collaboration, who provided spectroscopic and photometric catalogs used to select MOSDEF targets and to derive stellar population parameters. We acknowledge the First Carnegie Symposium in Honor of Leonard Searle for useful information and discussions that benefited this work.  
This research made use of Astropy,\footnote{http://www.astropy.org} a community-developed core Python package for Astronomy \citep{ast13, ast18}. Finally, we wish to extend special thanks to those of Hawaiian ancestry on whose sacred mountain we are privileged to be guests.

\section*{Data Availability}
The data underlying this article will be shared upon reasonable request to the corresponding author. The public MOSDEF data releases can also be found here: https://mosdef.astro.berkeley.edu/for-scientists/data-releases/ 

\vspace{5mm}
\noindent
\textit{Facilities}: \textit{Keck}/MOSFIRE, \textit{SDSS}

\noindent
\textit{Software}: Astropy \citep{ast13, ast18}, Corner \citep{for16}, FAST \citep{kri09}, IPython \citep{per07}, IRAF \citep{tod86, tod93}, Matplotlib \citep{hun07}, NumPy \citep{van11, harr20}, Pandas \citep{mckinney-proc-scipy-2010, pandas20}

\bibliographystyle{mnras}
\bibliography{references}

\bsp	
\label{lastpage}
\end{document}